\documentstyle[12pt,epsf]{article}

\renewcommand{\thefootnote}{\fnsymbol{footnote}}

\newcommand{\bea}{\begin{eqnarray}}
\newcommand{\eea}{\end{eqnarray}}
\newcommand{\beq}{\begin{equation}}
\newcommand{\eeq}{\end{equation}}

\begin{document}


\begin{titlepage}
\begin{flushright}
\begin{tabular}{l}
Preprint LPTHE - Orsay 98/33\\
Roma 1218/98\\
SNS/PH/1998-018
\end{tabular}
\end{flushright}
\vskip1.2cm
\begin{center}
  {\Large \bf Light hadron spectroscopy on the lattice with \\ the non-perturbatively  improved Wilson action}
  \vskip1cm 
{\large\bf D.~Becirevic$^a$, Ph.~Boucaud$^a$, L.~Giusti$^b$, J.P.~Leroy$^a$}\\
{\large\bf V.~Lubicz$^c$, G.~Martinelli$^d$, F.~Mescia$^d$, F.~Rapuano$^d$}\\

\vspace{.5cm}
{\normalsize {\sl $ ^a$Laboratoire de Physique Th\'eorique et Hautes Energies\\
Universit\'e de Paris XI, B\^atiment 211, 91405 Orsay Cedex, France \\
\vspace{.2cm}
$ ^b$ Scuola Normale Superiore and INFN,\\
Sezione di Pisa, P.zza dei Cavalieri 7, I-56100 Pisa, Italy\\
\vspace{.2cm}
$ ^c$ Dip. di Fisica, Univ. di Roma Tre and INFN,\\
Sezione di Roma Tre, Via della Vasca Navale 84, I-00146 Rome, Italy\\
\vspace{.2cm}
$ ^d$ Dip. di Fisica, Univ. ``La Sapienza" and INFN,\\
Sezione di Roma, P.le A. Moro, I-00185 Rome, Italy.}\\
  \vskip1.cm
  {\large\bf Abstract:\\[10pt]} \parbox[t]{\textwidth}{ 
We present results for the light meson masses and decay constants as obtained from calculations with the non-perturbatively improved (`{\sc Alpha}') action and operators on a $24^3\times 64$ lattice at $\beta = 6.2$, in the quenched approximation. The analysis was performed in a way consistent with ${\cal{O}}(a)$ improvement. We obtained: reasonable agreement with experiment for the hyperfine splitting; $f_K=156\pm 17\,{\rm MeV}$,
 $f_\pi =139\pm22\,{\rm MeV}$, $f_K/f_\pi = 1.13(4)$; $f_{K^*}=219\pm 7\,{\rm MeV}$
, $f_\rho =199\pm 15\,{\rm MeV}$, $f_\phi =235\pm 4\,{\rm MeV}$;  $f_{K^*}^{^T}(2\,{\rm GeV}) = 178\pm 10\,{\rm MeV}$, $f_\rho^{^T}(2\,{\rm GeV}) =165\pm 11\,{\rm MeV}$, where $f_V^{^T}$ is the coupling of the tensor current to the vector mesons; the chiral condensate $\langle\bar{q}q\rangle^{^{\overline{\rm MS}}} (2\,{\rm GeV})= - ( 253\pm 25\,{\rm MeV})^3$. Our results are compared to those obtained with the unimproved Wilson action. We also verified that the free-boson lattice dispersion relation describes our results very accurately for a large range of momenta.
\\
}}
\end{center}
\unboldmath
\end{titlepage}

\renewcommand{\thefootnote}{\arabic{footnote}}
\setcounter{footnote}{0}


\section{Introduction}
\setcounter{equation}{0}

From the very beginning of lattice QCD, one of the big challenges was to compute the hadron spectrum from first principles. In spite of the enormous technical progress that has been made in this field, yet there are ways to improve lattice studies systematically. The calculation of the light hadron spectrum is difficult mainly because of the large Compton wavelengths of the physical hadrons so that ever larger lattices are needed. On the other hand, to make a better contact with the continuum limit, simulations performed on several small lattice spacings are needed. These requirements are technically very demanding, and the search for systematic improvement is mandatory. 
Symanzik's proposal \cite{sym} for the improvement of the lattice action and quark bilinears with Wilson fermions, was realized perturbatively in \cite{sw,g2}. Fairly recently, the {\sc Alpha} collaboration {[4--10]} (see also \cite{a0}) has proposed and to a large extent carried out, a thoroughly non-perturbative method which aims the elimination of all  ${\cal{O}}(a)$ discretization errors. In this way, one of the most important sources of systematic uncertainties in numerical studies on the lattice, is practically removed.
The improvement program is implemented in several steps. The first source of ${\cal{O}}(a)$ uncertainties comes from the fermionic part of the Wilson lattice action. For on-shell quantities, these errors can be reduced to ${\cal{O}}(a^2)$, by adding one higher-dimensional operator only. The resulting action reads:
\bea
S_{SW} = S_{Wilson} + i c_{_{SW}} g_0  \left( {a^5 \over4} \sum_{x,\mu\nu} \bar q(x) \sigma_{\mu\nu} F_{\mu\nu}(x) q(x) 
\right).
\label{eq : swaction}
\eea
This is the Sheikholeslami-Wohlert or Clover action, where the last name is due to the shape of the lattice operator used for $F_{\mu\nu}(x)$. 
The non-perturbative determination of $c_{_{SW}}$, which is a function of the bare coupling only, allows the full non-perturbative improvement of the hadron spectrum. In Ref.~\cite{a2}, $c_{_{SW}}$ was determined non-perturbatively for different values of bare gauge coupling, and the final result of an overall fit for $g_0^2\leq 1$, is:
\bea
c_{_{SW}}(g_0)\, =\, {1 \,-\, 0.656\, g_0^2\, -\, 0.152\, g_0^4 \,-\, 0.054\, g_0^6 \over 1 \,- \,0.922\, g_0^2}.
\eea
For $\beta=6.2$ {\it i.e.} $g_0^2=6/\beta=0.9677$, this gives $c_{_{SW}}=1.614$, which is the value used in the present study{\footnote{In one-loop lattice perturbation theory at $\beta=6.2$, to one-loop order $c_{_{SW}}=1.257$, while with the so-called boosted coupling \cite{parisi,lpm}, $c_{_{SW}}=1.479$ (at tree level $c_{_{SW}}=1$). With these $c_{_{SW}}$-values, calculations were already performed several times (a recent review with a complete list of references can be found in \cite{rev}; see also \cite{rev98}).}}. 
The second source of  ${\cal{O}}(a)$ errors, comes from discretization effects in the matrix elements of composite local quark operators. These errors are relevant in the calculation of decay constants and/or form factors, {\it i.e.} quantities for which the knowledge of a hadronic ``wave function'' becomes crucial. As for the quark action, quark bilinears may also be improved through local counterterms, {\it i.e.} by adding specific operators which satisfy the same symmetry properties as the original ones. Most of the coefficients of the counterterms, which will be needed in this study, were calculated non-perturbatively. Their values will be given in course of the presentation.
Some of the counterterms are present only out of the chiral limit
and depend explicitly  on the quark masses. They come with   the so called $b$-coefficients. 
The only one  which is easy to obtain, from the forward matrix element of the
vector current, is  $b_V$,  which  has already been computed non-perturbatively~\cite{a1}.  
In order to determine $b$-coefficients for the other operators, the improvement 
program was extended in Ref.~\cite{g1} and most recently in \cite{gupta2}, but these proposals have not been applied yet to the {\sc ``Alpha''} action. Another attempt 
has been tried in Ref.~\cite{divitiis} but the results are not stable~\cite{dawson}. 
For these reasons, we have taken the values of all the other $b$-coefficients from perturbation theory. Their values, as well as those of the renormalization constants, will be quoted whenever used.

The implementation of the improvement  program in practical calculation was already done in \cite{gockeler} and in \cite{petronzio}. As for the heavy quark sector, only preliminary numbers have been reported so far \cite{ape&ukqcd}, and the final results will appear soon~\cite{prepa}. The main results of the present study, which concerns the light hadrons only, are given in the conclusion.

This paper is organized as follows: in Sect.~2 we give a short outline of the lattice setup, compute the spectrum of light mesons and extract $\kappa_{crit}$ in a way consistent with the improvement; in Sect.~3 we discuss the hyperfine splitting and the $J$-parameter; Sect.~4 is devoted to the study of decay constants; in Sect.~4, we also give our estimate for the chiral condensate; in Sect.~5, we make a comparison with  previous (unimproved) lattice results; in Sec.6 we test the energy-momentum relation on the lattice; we conclude in Sect.7.

\section{Hadron Masses, $\kappa_{critic}$ and $a^{-1}$}
\setcounter{equation}{0}

In this section, we will briefly discuss the standard procedure for extracting hadron masses and fix the lattice parameters ($\kappa_{critic}$ and $a^{-1}$). We will insist on details only when the procedure is different as compared to previous (standard) analyses.
     
\subsection{Lattice Setup and Hadronic Masses}
\setcounter{equation}{0}

Our results are based on a simulation performed on two Torre-APE100 (25 Gflops) machines located 
at the ``Roma - {\sc I}" University. Altogether, we have produced $100$ gauge field configurations on a lattice of 
size $24^3\times 64$ at $\beta = 6.2$, in the quenched approximation. 
After $5000$ Metropolis sweeps, obtained by starting from a cold configuration,
independent configurations were generated with a separation of 2000 sweeps. The values of the light Wilson hopping parameters used in our simulations, which are the same as in Ref.~\cite{gockeler}, are the following ones: 
\begin{itemize}
\item 0.1352 (u); 0.1349 (d); 0.1344 (s); 0.1333 (l).
\end{itemize}
The label assigned to the different quark masses (hopping parameters) are {\bf not} to be confused with actual quark masses. The quark propagators were inverted using the minimal residual algorithm preconditioned {\it \`a  la} Oyanagi~\cite{oyanagi}. 

To estimate the statistical errors, the raw results for various correlators (see below), as obtained on our 100 configurations, were jackknifed by decimating five of them at the time{\footnote{We also tested that by varying the number of configurations per cluster, the error estimates remain stable.}.} 

Enabled by our correlators, we proceed the analysis by plotting the effective masses for pseudoscalar and vector mesons.  
\begin{figure}[t!]
\vspace*{-2.cm}
\begin{center}
\begin{tabular}{@{\hspace{-2.1cm}}c c c}
$$\epsfbox{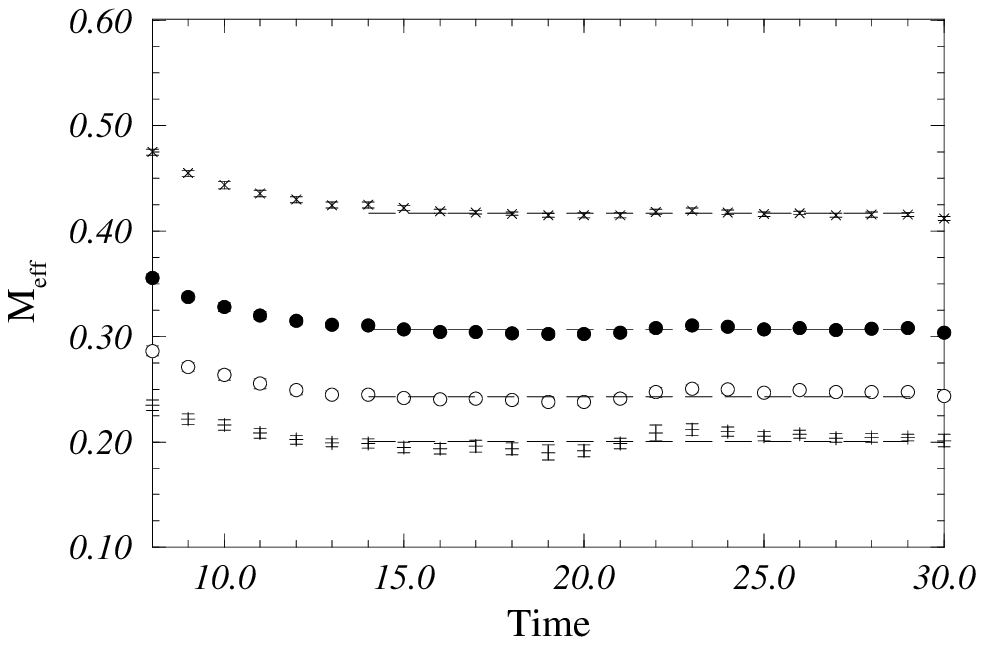}$$ &\hspace*{-0.5cm} & $$\epsfbox{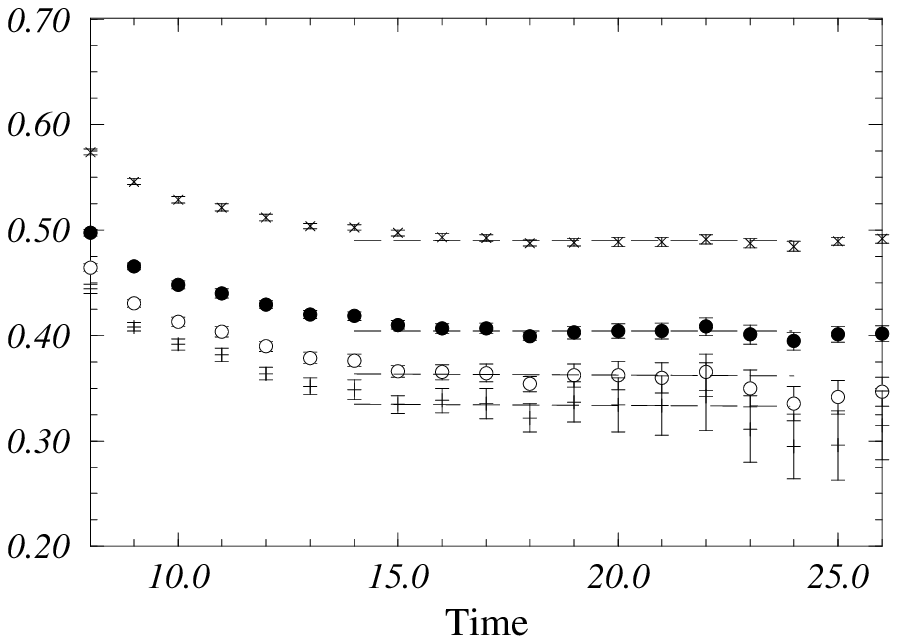}$$ \vspace*{.1cm}\\
\end{tabular}
\caption{\it Effective masses for pseudoscalar (left - ) and vector (right figure) mesons. From up to down, four curves in each figure correspond to mesons containing {\it `ll', `ss', `dd', `uu'} quark flavours, respectively.}
\label{fig : plot0}
\end{center}
\end{figure}
These plots are shown in Fig.~\ref{fig : plot0}, where the effective masses are obtained as solutions of:
\bea
\frac{C_{_{JJ}}(t+1)}{C_{_{JJ}}(t)} = {\rm cosh} M_{J}\,\left( 1 \,-\, {\rm tanh}M_{J}\,{\rm tanh}\left[ M_{J} \left( {T\over 2} - t \right) \right] \right).
\label{eq : meff1}
\eea
The hadronic masses in lattice units will be denoted by $M_J$, whereas the physical ones by $m_J$, {\it i.e.} $M_J=a m_J$. Time distances and space coordinates are always expressed in lattice units.
By $C_{_{JJ}}(t)$, we generically refer to the usual $two-point$ correlation function summed over $\vec x$, which (for large euclidean times) is dominated by the lightest hadronic state which couples to the chosen interpolating current $J$:
\bea
C_{_{JJ}}(t) = \sum_{\vec x} \langle 0 \vert J({\vec x}, t) J^{\dagger}(0)\vert 0 \rangle\, &\stackrel{t\gg 0}{\longrightarrow}&\,  {{\cal{Z}}_J \over 2 M_J} \left( e^{- M_J t} + \eta \,e^{- M_J (T - t)} \right) \nonumber \\
&\,=\,& {{\cal{Z}}_J \over  M_J}e^{- M_J T/2} {\rm cosh}\left[ M_J \left( {T\over 2} - t \right) \right] .
\label{eq : meff}
\eea
$T=64$ is the lattice temporal extension, and $\eta$ the temporal inversion ${(t \leftrightarrow T - t)}$  symmetry factor, which is $+1$ in the $JJ$-case for mesons{\footnote{Among the correlators considered in this study, only in the case $C_{_{AP}}(t)$ (corresponding to the correlator of the fourth component of the axial current  with the pseudoscalar density), one has: $\eta=(-1)$.}}. Note that, this (`cosh') form has been used to obtain the relation (\ref{eq : meff1}). For pseudoscalar mesons, the standard interpolating current that couples to the pion ($J^{PC}=0^{-+}$) is ${J_{PS}(x)=i\bar{q}(x) \gamma_5 q(x)}$. For the extraction of quantities related to the $\rho$-meson ($J^{PC}=1^{--}$), the local vector current,
${J_{\mu}(x)=\bar{q}(x)\gamma_\mu q(x)}$, is the appropriate one.   More specifically, we consider the space component $J_{i}(x)$ and average over the indices, which is the procedure usually employed to reduce the statistical noise. By using effective mass plots, we may fix the initial time ($t_{in}$) of the range on which we fit the data to extract the lightest masses{\footnote{At that $t_{in}$, we assume that contributions of higher excitations which couple to a given correlation function are negligible.}}. The final time is best fixed by direct inspection of the signal to noise ratio in the hadronic propagator $C_{_{JJ}}(t)$. With these two criteria, we establish the fit intervals: our light pseudoscalar mesons are well isolated for $t \in [14, 29]$
, while the vector ones for  $t \in [14, 24]$.
\noindent

From the fit~(\ref{eq : meff}), we obtain the hadronic masses in lattice units $M_J$. The fit parameters are reported in Tab.~\ref{tab : mass}, where we also present the results of our extrapolation to $\kappa_{crit}$  which we discuss now.


\begin{table}[h!]
\begin{center}
\begin{tabular}{|c|c|c|c|c|} \hline
{\phantom{\Huge{l}}}\raisebox{-.1cm}{\phantom{\Huge{j}}}
{ ``flavor''} & { $M_{PS}$}  & { ${\cal{Z}}_{PS}$} & { $M_{V}$}  & { ${\cal{Z}}_V$} \\ \hline \hline
{ $\ell \ell$} & 0.4167(15) & 0.0111(4) & 0.4911(29) & 0.0037(2) \\
{ $ss$} & 0.3058(19) & 0.0077(4) & 0.4055(47) & 0.0022(2) \\
{ $dd$} & 0.2440(21) & 0.0063(4) & 0.3626(78) & 0.0016(2) \\
{ $uu$} & 0.2007(26) & 0.0057(4) & 0.335(12) & 0.0013(3)\\ \hline
{ critical}& -- & 0.0035(4) & 0.275(22) & 0.0005(3) \\ \hline 
\end{tabular}
\vspace*{.2cm}

\caption{\it Masses and {${\cal{Z}}$}'s for pseudoscalar and vector mesons in lattice units. These results are in good agreement with results of {\rm Ref.~\cite{gockeler}}.}
\label{tab : mass}
\end{center}
\end{table}

\vspace*{.1cm}

\subsection{Critical Parameter and Inverse Lattice Spacing}

\begin{figure}
\vspace*{-1.1cm}
\begin{center}
\begin{tabular}{@{\hspace{-0.7cm}}c c c}
  &$$\epsfbox{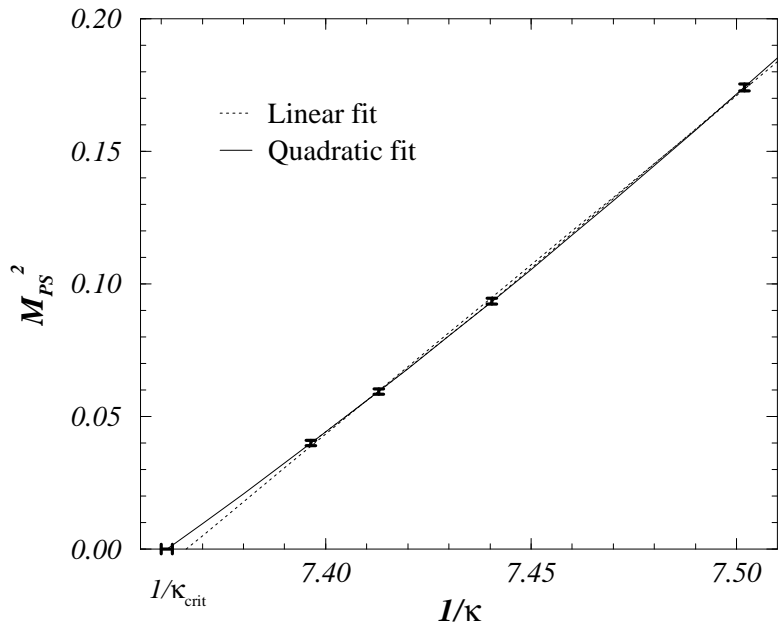}$$ & \\
\end{tabular}
\vspace*{-.9cm}
\caption{\it Quadratic vs. linear extrapolation of $M_{_{PS}}^2$ to $\kappa_{crit}$.}
\label{plot2}
\vspace*{.3cm}
\end{center}
\end{figure}

We now discuss the uncertainties in the determination of $\kappa_{crit}$ which depend on the method used to fit the pseudoscalar meson masses. $\kappa_{crit}$ represents the value at which the chiral symmetry on the lattice should be restored. In practical calculations, the basic relation is the Gell-Mann-Oakes-Renner one \cite{gmor}:
\bea
m_\pi^2 = - {4\over f_\pi^2} \langle \bar q q\rangle  \, m_q   \, +\, {\cal{O}}(m_q^2)~,
\label{eq:aa}
\eea
which states that the terms responsible for explicit chiral symmetry breaking are linear in the quark masses.
On the lattice, for degenerate quark masses and by neglecting the terms of order ${\cal{O}}(m_q^2)$, this implies:
\bea
M_{PS}^2 = \alpha m_q = {\alpha\over 2} \left( {1 \over \kappa_{q}} - {1 \over \kappa_{crit}} \right), 
\label{eq:aa1}
\eea
where the standard definition ({\sl that can be derived from the vector Ward identity}) of $m_q$ has been employed:
\bea
a m_q \,=\, {1\over 2} \left( {1\over \kappa_q} - {1 \over \kappa_{crit}}\right).
\label{eq:qmass1}
\eea
However, this ({\it standard}) procedure to determine $\kappa_{crit}$ is valid if we have points sufficiently close to the chiral limit so that, up to chiral logarithms, higher-order quark mass terms can be neglected. These terms can arise from two sources. On the one hand, they are due to the lattice artifacts  and can be eliminated by replacing $a m_q \, \to \, am_q ( 1 \,+\, b_m \,am_q)$. On the other, they can be a real physical effect, as indicated in (\ref{eq:aa}). In order to investigate this point, we made a fit of the form:
\bea
M_{PS}^2\, = \, \alpha_1 \, \left( {1\over \kappa_q} -  {1\over \kappa_{crit}}\right)\,+\, \alpha_2 \, \left( {1
\over \kappa_q} -  {1\over \kappa_{crit}}\right)^2 
\label{eq1}
\eea
from which we obtained our best estimate:
\bea
\bullet\quad \kappa_{crit}^{quad}= 0.135845(25)
\label{eq1bis}
\eea
with $\alpha_1 = 1.106(32)$, $\alpha_2 = 0.94(13)$. From the result of the fit, and as can be seen in Fig.~\ref{plot2}, the sign of $\alpha_2$ is opposite to what one would expect from the present determination of $b_m$. For example, in lattice (boosted) perturbation theory  $b_m = - 0.593$ ($b_m = - 0.652$)~\cite{a7} {\footnote{ A non-perturbative estimate, $b_m = - 0.62(3)$, was given in \cite{divitiis}. We tried to use the same technique but the values that we obtain are very unstable.}}. Thus, unless perturbation theory gives the opposite sign (which we believe it is impossible), this implies that the positive curvature is {\it a physical effect}. The value of $\kappa_{crit}$ (\ref{eq1bis}) is the one that we will use throughout this and our forthcoming studies. We note, in passing, that the result of a linear fit with the three lighter mesons gives $\kappa_{crit}= 0.135801(19)$. 
\noindent
An important observation is that ~(\ref{eq1bis}) agrees with the value $\kappa_{crit}$ obtained from the axial Ward identity \cite{maiani}: 
\bea
{\rho}_{_{WI}}\, = \,{{\langle \partial_0 {{A}}_0(t) {\cal{O}}^{\dagger}(0)\rangle} \over { 2 \,\langle {{P}}(t) {\cal{O}}^{\dagger}(0)\rangle}}\, +\, {\cal{O}}(a^2).
\label{eq:qmass22}
\eea
\noindent
In the above formula, the axial current was improved as $A_\mu(x)\to A_\mu(x) + c_A \partial_\mu P(x)$. A hat denotes that the quantity is properly renormalized  {({\sl see Sect.4})}. With the simple choice of ${\cal{O}}(t) = P(t)$, and by using the symmetric definition of the lattice derivative, at large time distances,  we practically fit the following:
\bea
\rho_{_{WI}} &=& \rho^{(0)} + c_{_A} a \rho^{(1)}{\quad \rm where }\nonumber \\
\rho^{(0)} &=& {\langle A_0(t) P(0)\rangle \over 2\, \langle P(t) P(0)\rangle}\, {\rm sinh}\left( M_{PS}\right) \nonumber \\
a \rho^{(1)} &=&  {\rm cosh}\left( M_{PS}\right) - 1 
\eea
Results are presented in Tab.~\ref{tab:100}.

\begin{table}[h!]
\begin{center}
\begin{tabular}{|c|c|c|c|} \hline 
{\phantom{\Huge{l}}}\raisebox{-.1cm}{\phantom{\Huge{j}}}
{ $\kappa_1\,\kappa_2$}& { $\rho^{(0)}$} & { $- c_{_A}\, {a \rho^{(1)} / \rho^{(0)}}$} &  { $\rho_{_{WI}}$}  \\ \hline \hline 
{ $\ell \ell$} & 0.0794(15) & 0.0410(8) & 0.0764(15) \\

{ $ss$} & 0.0450(15) & 0.0388(12) & 0.0432(15)  \\

{ $dd$} & 0.0292(16) & 0.0378(19) & 0.0281(16) \\

{ $uu$} & 0.0199(16) & 0.0373(29) & 0.0192(17) \\ \hline 
\end{tabular}
\label{tab:100}

\vspace*{.3cm}

{\caption{{\it Lattice axial WI quark bare masses $\rho$'s from mesons consisted of degenerate `flavors'. The results are given in lattice units. Note that the} ${\cal{O}}(a)$ {\it correction $c_{_A} a \rho^{_{(1)}}/ \rho^{_{(0)}}$ never exceeds $\sim 4\%$.}}}
\end{center}
\end{table}
%
\noindent
The resulting value for $\kappa_{crit}$ extracted from the linear dependence, $\rho(1/\kappa)$, is{\footnote{We checked that the result of the quadratic fit is indistingishable from the value we quote, $\kappa_{crit}^{^{AWI}}$.}}: 
\bea
\bullet \quad \kappa_{crit}^{^{AWI}}= 0.135840(48) \,,
\label{wiwi}
\eea
\noindent
in very good agreement with (\ref{eq1bis}). Note also that (\ref{wiwi}) agrees with $\kappa_{crit}=0.135828(5)$, obtained by the same method in \cite{petronzio}, as well as with the result of Ref.~\cite{gockeler}: {$\kappa_{crit}\,=\,0.13589(2)$. \\
Evidently, these estimates differ sensibly (as expected), when compared to the (boosted) perturbative value, at $\beta=6.2$:  $\kappa_{crit}^{BPT}\,=\, 0.1374$ \cite{g2,a4}.  \\

We now discuss the calibration of the inverse lattice spacing.
A conventional method to set the scale is obtained by extrapolating the vector meson mass to the chiral limit and compare the obtained result in lattice units, $M_\rho$ (see Tab.1.), to the experimental value {$m_\rho = 768\,{\rm MeV}$}.
Using a quadratic fit of $M_V$ in the quark masses, we obtain:
$a^{-1}=2.69(14)\,{\rm GeV}$
.
This value is in good agreement with our preferred value for $a^{-1}=2.75(17)\,{\rm GeV}$ that we discuss in the next subsection. 
In the real world {\sl (unquenched QCD)}, to fix the physical scale, we can use any hadronic quantity {\it e.g.} $f_\pi, m_p, ...${\footnote{In that case we would not use $m_\rho$, since the $\rho$-meson is not a stable physical particle.}}
In the quenched approximation however, different physical quantities can lead to different calibrations of the lattice spacing. The calibration of $a^{-1}$ from different quantities, ($m_\rho$, $m_{K^*}$, $f_\pi$, $f_\rho$ etc.), differ by less than the quoted statistical errors. For this reason we are unable to study this systematic effect.

\subsection{`Lattice physical planes' - procedure}

The uncertainty due to the extrapolation to $\kappa_{crit}$ can be circumvent if we adopt the so called {\bf ``method of lattice physical planes''}, which was proposed and used in Ref.~\cite{giusti}. One starts with a definition of two physical lattice planes, {\it i.e.} expresses  $M_V$ and some other physical quantity obtained by lattice calculations, generically denoted by $\Phi$, as functions of $M_{PS}^2$. In the plane $(M_V,M_{PS}^2)$, one looks for the point where a fit to the lattice data meets the curve $M_V = C_{s\ell}\sqrt{M_{PS}^2}$, with $C_{s\ell}$ fixed by its experimental value, $m_{K^*}/m_K$. The point where the two curves cross determines $M_K$ and $M_{K^*}$. At the same value of $M_K^2$ in the plane $(\Phi,M_{PS}^2)$, one reads off the corresponding $\Phi_{K,K^*}$, in the kaon sector. Similarly, by fixing $C_{\ell \ell}= m_{\rho}/m_\pi$,  we can obtain $\Phi_{\pi,\rho}$, without direct extrapolation to $\kappa_{crit}$. In this study, all our data are fitted quadratically in $M_{PS}^2$. 

To fix the value of the inverse lattice spacing, one proceeds as follows. From the cross in the first plane $(M_V,M_{PS}^2)$, of fitted data with the curves $M_V = C \sqrt{M_{PS}^2}$ (with $C=C_{s\ell}$ or $C=C_{\ell\ell}$), we get:
\bea
M_\pi = 0.0491(42),\quad {\rm and}\quad M_\rho = 0.279(24);\nonumber \\
M_{K} = 0.180(12),\quad {\rm and}\quad M_{K^*} = 0.321(21).
\label{plane0}
\eea
Using $M_\rho$, we fix the scale as $a^{-1}=m_\rho/M_\rho$ and obtain: $a^{-1}(m_\rho)=2.75(22)\, {\rm GeV}$ . In \cite{giusti}, it was argued that due to the fact that the mass of ${K^*}$ is in range of masses directly accessible on the lattice, $m_{K^*}$ is the most suitable quantity for the scale fixing. We adopt this argument and get:
\bea
 a^{-1}(m_{K^*})=2.75(17)\, {\rm GeV}, 
\eea
\noindent
which is the value that we will use in the following discussion and all our forthcoming studies.

By using the same method, one can also determine the value of the light quark mass. Details of this analysis were presented in Ref.~\cite{vittorio}. Here, we only give the value of $\kappa_{str}$, {\it i.e.} the one which corresponds to the strange quark mass:
\bea
\bullet \quad \kappa_{str}= 0.13482(12) \,.
\eea
By using this result in Eq.~\ref{eq1}, we obtain the hypothetical pseudoscalar: $M_{\eta_{\bar{s}s}}^2=0.0646(75)$. Then, in the plane $(M_V,M_{PS}^2)$, we read-off:
\bea
M_{\phi}=0.369(18)\quad \to \quad m_{\phi}=1.013(19)\, {\rm GeV}.
\eea
where we used $a^{-1}(m_{K^*})$.  

\section{Hyperfine Splitting}
\setcounter{equation}{0}

One of the main problems of lattice studies with Wilson fermions
is to reproduce the experimentally observed, {\sl approximately constant} hyperfine splitting. Theoretically, one expects {$m_V^2 - m_{PS}^2$} to be constant in the heavy quark limit only. Experimentally, one finds:\\

\vspace*{-3mm}
$(m_{D^*}^2 - m_{D}^2)^{(exp)} = 0.546\, {\rm GeV}^2$,\hspace*{3mm}    $(m_{B^*}^2 - m_{B}^2)^{(exp)} = 0.485\, {\rm GeV}^2$ . \vspace*{2mm} \\
There is no theoretical reason to explain why the hyperfine splitting has almost the same value also for light mesons~ \cite{pdg}:\\

\vspace*{-3mm}
$(m_{\rho}^2 - m_{\pi}^2)^{(exp)} = 0.571\, {\rm GeV}^2$,\hspace*{3mm}   $(m_{K^*}^2 - m_{K}^2)^{(exp)} = 0.552\, {\rm GeV}^2$.  \vspace*{2mm} \\
The net effect of the Clover term in the improved action is to give an extra anomalous magnetic moment to quarks, which increases and flattens the lattice hyperfine splitting. The most convenient method to study the splitting is to extract $M_V - M_{PS}$ directly from the ratio:
\bea
{C_{_{VV}}\over C_{_{PP}}}(t) \, \sim \, e^{-(M_{_V}-M_{_{PS}}) \,t} .
\eea  
\\
The results for the mass difference and the hyperfine splitting are listed in  Tab.~\ref{tab:split}, and displayed in Fig.~\ref{fig : plot1}. 
{
\begin{table}[h!]
\begin{center}
\begin{tabular}{|c|c|c|} \hline
{\phantom{\Huge{l}}}\raisebox{-.1cm}{\phantom{\Huge{j}}}
{ ``flavor''}&  { $M_{V}-M_{PS}$}  &{ $M_{V}^2-M_{PS}^2$}  \\ \hline \hline
{ $\ell \ell$} &  0.0735(22) & 0.0668(22)  \\
{ $ss$} &  0.1006(42) & 0.0716(34)  \\
{ $dd$} &  0.1219(83) & 0.0740(59) \\
{ $uu$} &  0.1398(140) & 0.0749(90) \\ \hline
\end{tabular}
\vspace*{.3cm}
\caption{\it Mass splitting in lattice units.}
\label{tab:split}
\end{center}
\end{table}
\begin{figure}[h!]
\begin{center}
\begin{tabular}{@{\hspace{-0.7cm}}c c c}
  &$$\epsfbox{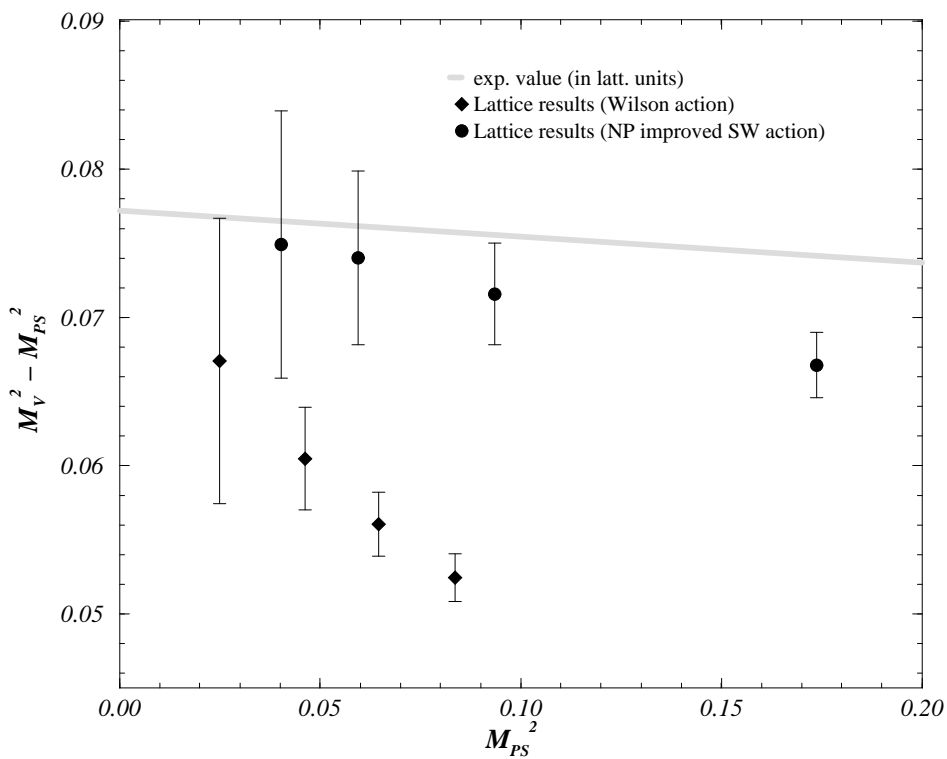}$$ & \\
\end{tabular}
{\caption{\it Hyperfine splitting on the lattice. The experimental line of results is given in lattice units by assuming $a^{-1}(\sigma)=2.72(3)\,{\rm GeV}$.}}	
\label{fig : plot1}
\end{center}
\end{figure}}
\noindent
\hspace*{-3mm} In this figure, we also show the results obtained with unimproved Wilson fermions (at the same $\beta=6.2$ and on the same volume) \cite{giusti}. For comparison with experiment, we translated the experimental $m_V^2 - m_{PS}^2$, into lattice units. To this purpose, we take $a^{-1}(\sigma)=2.72(3)\,{\rm GeV}$, extracted from the computation of string tension \cite{bali}. We observe that the data obtained using the Clover action describe much better the experimental hyperfine splitting. 

It is encouraging that the Clover term reduces the slope in $M_{PS}^2$ which characterizes the Wilson data. Moreover, the values are much closer to the experimental ones.  
Our preliminary study of the heavy hadron spectrum \cite{prepa}, however, shows that for heavy-light meson, all our results for the hyperfine splitting are well below the experimental values. 

A consistent comparison between results for several physical quantities obtained with improved and the Wilson actions, including the hyperfine splitting, will be reported in Sect.~5.

A frequently used quantity to test the effects of quenching (or other systematic errors) is the dimensionless, so-called $J$-parameter \cite{maiani, michael}:
\bea
J=M_{V} { d M_{V} \over d M_{PS}^2} .
\eea
In the kaon-sector, results obtained with the Wilson action are always below the experimental value, $J^{(exp)}=0.48$. By fitting the $M_V$ linearly in $M_{PS}^2$, the value for $J$ that we obtain is:
\bea
J=0.373(7), 
\eea
which is to be compared to $J=0.36(2)$ and $J=0.34(5)$ \cite{giusti}, obtained on the same lattice and at the same $\beta=6.2$, but with Wilson and tree-level improved ($c_{_{SW}}=1$) Clover action, respectively.
When we fit quadratically: $M_V = {\cal{A}} M_{PS}^4 + {\cal{B}} M_{PS}^2 + {\cal{C}} $, we get:
\bea
J= M_{K^*} ({\cal{B}} + 2 {\cal{A}} M_{K}^2 )= 0.47(6).
\eea
with ${\cal{A}}=-1.91(98)${\footnote{Note a small difference of ${\cal{A}}$ and the value reported in \cite{vittorio}, which is due to the shorter time interval for the fit of $C_{VV}$ that we use in this study. This difference is however irrelevant for our final results.}}, ${\cal{B}} = 1.57(31)$, and ${\cal{C}} =0.275(22)$.
Of course, the present statistical uncertainties, as well as the small number of hopping parameters that we use in this study, do not allow for a definite conclusion on this issue. In particular, the mesons with nondegenerate flavors are important to establish better the quadratic coefficient ${\cal{A}}$, since the terms proportional to the square of the difference of the quark masses may give some contribution as well. Some more research to further investigate this point, is needed.

\section{Decay constants}
\setcounter{equation}{0}

With the use of non-perturbatively renormalized improved operators, $f_\pi$ becomes an equally good candidate as the hadron masses for fixing the inverse lattice spacing - $a^{-1}$.

The standard procedure to extract the pseudoscalar and vector decay constants consists in calculating the following quantities:
\bea
{C_{AP}(t)\over C_{PP}(t)}\,=\,
{{\sum_{\vec x} \langle {\hat{A}_0}(\vec x,t) {\hat{P}}(0)\rangle} \over 
{\sum_{\vec x} \langle {\hat{P}}(\vec x,t) {\hat{P}}(0)\rangle}}  &\simeq&  {\hat{F}_{PS}} \, {M_{PS}\over {\sqrt{{\cal{Z}}_{PS}}}}\, {\rm  tanh} \left(M_P (\frac{T}{2} - t)\right) \label{raspad} \\
\vspace*{-.6cm} 
C_{VV}(t)\,=\, \sum_{\vec x} \langle {\hat{V}_i}(\vec x,t) {\hat{V}_i}(0) \rangle   &\simeq&  M_V^2 {\hat{F}_V}^2 e^{-M_V {T \over 2}}\, {\rm  cosh} \left(M_V (\frac{T}{2} - t)\right)
\label{raspad1}
\eea
where the `cosh'-form of fit (\ref{eq : meff}) is assumed{\footnote{In the case of the pseudoscalar decay constant, we could also use $C_{A A}(t)$. The reason why we use the one in Eq.~(\ref{raspad}), is that in this case the errors on decay constant are smaller.}}. Note that we consistently use capital letters to emphasize that the quantity is given in lattice units; to distinguish it from the corresponding counterpart in physical units, which is denoted by lower-case letters. $\hat{F}_{i}$ refer to the decay constants of currents which include suitable overall renormalization constants.
 The pseudoscalar, ${\hat{F}_{PS}}$, and the vector, ${\hat{F}_V}$, decay constants are defined as usual:
\bea
\langle 0|{\hat{A}_0}|PS(\vec p = 0)\rangle &=& i {\hat{F}_{PS}} M_{PS}\hspace*{6mm} {\rm{and}}\nonumber \\
\langle 0|{\hat{V}}_i|V(\vec p = 0)\rangle &=& i e_i^{(\lambda)} {\hat{F}_V} M_V 
\eea
The improvement of the axial and vector currents is achieved by adding the derivative of the pseudoscalar density, $\partial_\mu P$, and the divergence of the tensor current, $\partial_\nu T_{\mu \nu}$, respectively:
\bea
\langle 0|{\hat{A}}_0|PS(\vec p = 0)\rangle &\to& Z_A\,(1 + b_{A} a m_q ) \,\left[ \langle 0|A_0|PS\rangle + c_A \langle 0|a \partial_0 P|PS\rangle \right] \nonumber \\
& & =\, i M_{PS} {\hat{F}_{PS}} \, =\, i M_{PS} ({\hat{F}_{PS}}^{(0)} + c_A a {\hat{F}_{PS}}^{(1)})\\
\langle 0|{\hat{V}}_i|V(\vec p = 0)\rangle &\to& Z_V\,(1 + b_{V} a m_q ) \,\left[ \langle   0|V_i|V\rangle + c_V  \langle   0|a \partial_0 T_{i0}|V\rangle \right] \nonumber \\
& &=\, i M_V e_i^{(\lambda)} {\hat{F}_V}\, =\, i M_V e_i^{(\lambda)} ({\hat{F}_V}^{(0)}+ c_V a {\hat{F}_V}^{(1)})
\eea
where for clarity, $a$ in corrective terms on the {\it r.h.s.} are written explicitly. $c_V$ and $c_A$ are suitable coefficients, the values of which are chosen as to cancel ${\cal{O}}(a)$ errors in physical matrix elements; the renormalization constant $Z_{A}$, $b_{A}$, $Z_{V}$ and $b_{V}$ will be discussed later on. By observing that:
\bea
{\langle \partial_0 P(t) P(0)\rangle \over \langle P(t) P(0)\rangle} = - {\rm  sinh}( M_{PS} ),
\eea
the corrections to the decay constants in actual calculations are then obtained using:
\beq
a F_{PS}^{(1)} =  {{\sqrt {{\cal{Z}}_{PS}}}\over M_{PS}}  {\rm  sinh} (M_{PS})\ \hspace*{6mm} {\rm{and\, similarly,}}\hspace*{6mm}
a F_V^{(1)} =  {{\sqrt {{\cal{Z}}_{V}}}\over M_V}  {\rm  sinh}(M_V).
\eeq
With all these relations, we extract the decay constants and list their values in Tab.~\ref{tab : ctst}.

\begin{table}
\begin{center}
\begin{tabular}{|c|c|c|c|c|c|c|} \hline 
{\phantom{\Huge{l}}}\raisebox{-.1cm}{\phantom{\Huge{j}}}
{ $\kappa_1\,\kappa_2$}& { $F_{PS}^{(0)}$} & { $a F_{PS}^{(1)}$} &  { $F_{PS}$} & {  $F_V^{(0)}$} & { $a F_V^{(1)}$} &  { $F_V$} \\ \hline \hline
{ ${\ell \ell}$} & 0.0939(21) & 0.1090(19) & 0.0899(21) & 0.1235(24) & 0.0460(12) & 0.1136(23)\\

{ $ss$} & 0.0832(27) & 0.0891(21) & 0.0799(27) & 0.1161(32) & 0.0349(14) & 0.1086(32)  \\

{ $dd$} & 0.0770(36) & 0.0798(20) & 0.0740(35)& 0.1105(47) & 0.0305(24) & 0.1039(47) \\

{ $uu$} & 0.0733(47) & 0.0754(21) & 0.0706(46) & 0.1061(69) & 0.0303(46) & 0.0996(70) \\ \hline

{ critic} & 0.0675(47) & 0.0636(23) & 0.0652(47) & 0.0996(88) & 0.0221(58) & 0.0949(87)   \\ \hline 

\end{tabular}
\caption{\it Decay constants -- in lattice units.}
\label{tab : ctst}
\end{center}
\end{table}
%

\noindent
In that table, the improvement coefficient ${c_A=-0.037}$ was used. It was obtained in \cite{a2} with an overall fit at $g_0^2 \leq 1$:
\bea
c_A= -0.00756\, g_0^2\, {1\, - \,0.748\, g_0^2\over 1\, - \,0.977\, g_0^2}\,.
\eea
We also take $c_V=-0.214$, as suggested by the preliminary study of \cite{a8}. Here, we did not account for the errors they quote, {\it i.e.} $c_V=-0.214(74)$. In the case of light quarks, the improvement term ($aF_V^{(1)}$) is {\it very small} anyway. Still, one comment is in order. Of all improvement coefficients, only $c_V$ differs by one order of magnitude from its (boosted) perturbative value $c_V=-0.026$. For this reason, to be totally on the safe side, we have also calculated $F_V$ with the perturbative $c_V$ given above. In this case the results for vector decay constants that we present in this paper are increased by about $5\%$. While this effect is less pronounced in the light hadron case (since the correction proportional to $c_{_V}$ is rather small), it turns out to be very important in determination of the heavy-light vector meson decay constants \cite{prepa}.

We now discuss the values of the renormalization constants
which have been used to obtain the physical currents $\hat{V}_\mu$ and $\hat{A}_\mu$.

The renormalization constants, $Z_V$ and $Z_A$, depend on the external quark masses, {\it i.e.} $Z_{A,V} = Z_{A,V}(m)$ . In the chiral limit, both constants were calculated non-perturbatively in \cite{a1}, with the following results:
\bea
Z_V&=& {1\, - \,0.7663\, g_0^2\, +\, 0.0488\, g_0^4 \over 1\, -\, 0.6369\, g_0^2} \cr
Z_A&=& {1 \,-\, 0.8496\, g_0^2\, +\, 0.0610\, g_0^4 \over 1\, -\, 0.7332\, g_0^2} .
\eea
In our case, it gives $Z_A\,=\,0.8089$, and $Z_V \,=\, 0.7927$.  
The last step in relating a lattice decay constant to its continuum value is to account for the explicit quark  mass corrections. The constant $b_V$ has been computed non-perturbatively. The global fit, first given in \cite{a1}, was updated in \cite{a7}:
\bea
b_V = {1\, -\, 0.7613 \,g_0^2\, + \,0.0012\, g_0^4\, -\, 0.1136 \, g_0^6 \over 1 \, - \, 0.9145 \, g_0^2}
\eea
giving in our case: $b_V = 1.404$. The other constants are not known non-perturbatively and we have to rely on perturbation theory.
To one loop accuracy, we have: $b_J = 1 + b_J^{(1)} g_0^2$, with
$b_V^{(1)}\,=\,C_f \,0.11492(4),\, b_A^{(1)}\,=\,C_f\, 0.11414(4)\cite{a7},\, b_T^{(1)} \,=\, C_f\, 0.10434(4)\cite{a6}$, with {${C_f = (N^2 - 1) / 2 N}$}.\\
Boosted perturbative values are obtained by the replacement  $g_0 \to g_B$\footnote{$g_B^2 = g_0^2 / \langle P\rangle$ where the average plaquette $\langle P\rangle = 0.6136$ for $\beta=6.2$ as inferred by our Monte Carlo.}. For comparison, in this way we get $b_V = 1.242$, rather close to the non-perturbative result. This makes us confident that by using for the constant $b_A$, the result of boosted perturbation theory, $b_A=1.240$, we committed negligible error for the light meson decay constants considered here.
The corrective coefficients $b_{V,A}$ enter with the bare quark mass $am_q$ defined in (\ref{eq:qmass1}).

In order to extract the desired physical quantities, we are supposed to make an extrapolation to the chiral limit as well as an interpolation to the strange mass sector. The results of a quadratic extrapolation to $\kappa_{crit}$ are given in Tab.~\ref{tab : ctst}{\footnote{ We checked that the extrapolation of decay constants remains the same regardless of whether we combine $\hat{F}^{(0,1)}_i$ before or after extrapolation. For heavy mesons, however, $\hat{F}^{(0)}_i$ and $\hat{F}^{(1)}_i$ must be combined before extrapolation \cite{prepa}.}}.
This procedure is basically equivalent to the case of massless pion in the method of physical lattice planes. 
The method of lattice planes, as applied to our data, is illustrated in Fig.~\ref{plane}.
\begin{figure}[t!]
\vspace*{-2.cm}
\begin{center}
\begin{tabular}{@{\hspace{-1.8cm}}c c c}
$$\epsfbox{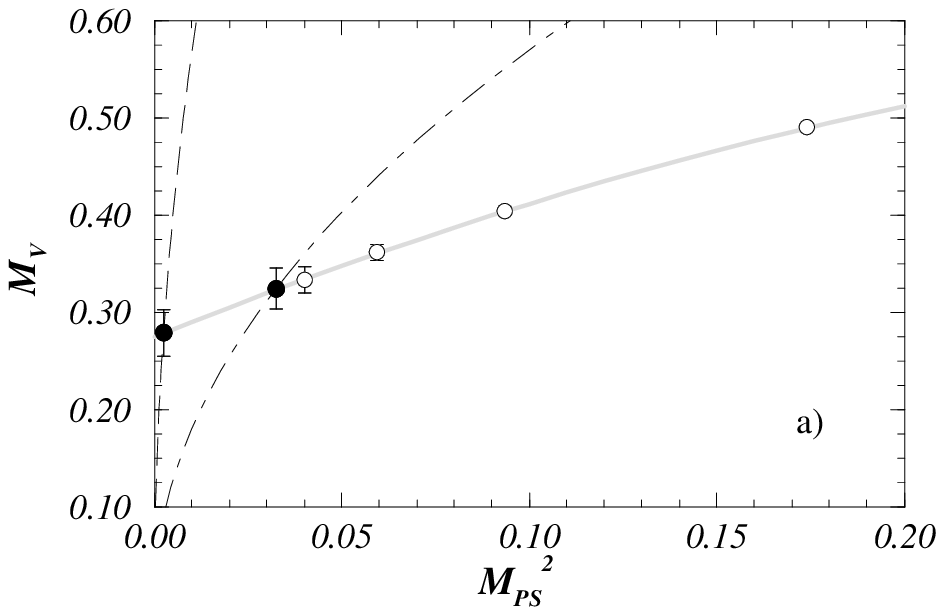}$$ &\hspace*{-3.3cm} & $$\epsfbox{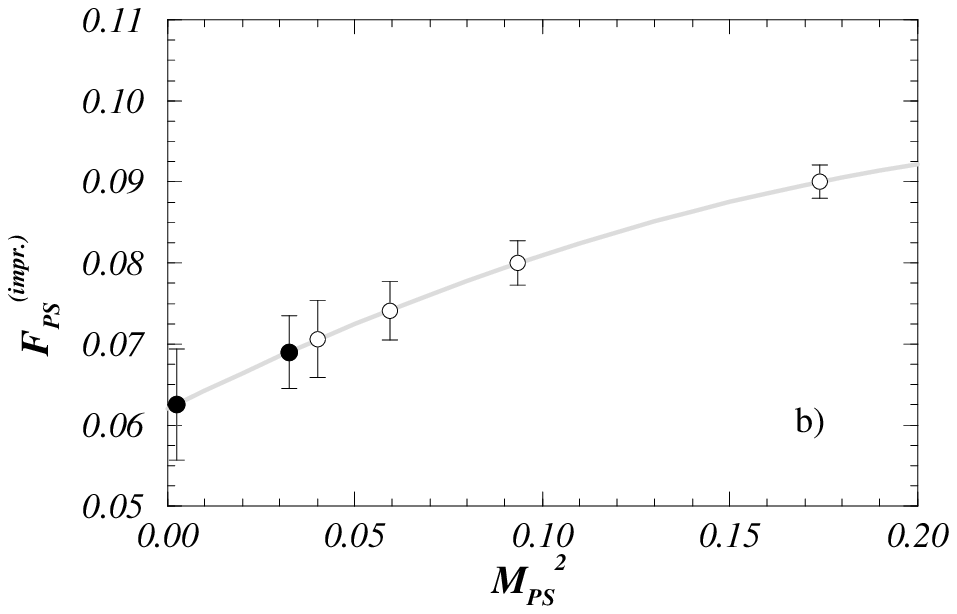}$$ \vspace*{.6cm}\\
\end{tabular}
\caption{\it Method of planes: In the plane a), dot-dashed line is obtained by fixing $C_{s\ell}$ while the dashed one by fixing $C_{\ell \ell}$. Cross-points with the line which quadratically fits the data, fix $M_K^2$ and $M_\pi^2$ at which one reads off the values of the decay constant in the plane b), where the quadratic fit has also been used.}
\label{plane}
\end{center}
\end{figure}
\begin{itemize}
\item  In the plane $(F_{PS},M_{PS}^2)$, we extract {\underline{the pseudoscalar decay constants}}, using $M_\pi = 0.0491(42)$ and $M_{K} =  0.1801(115)$ as determined with the lattice planes method (see Eq.~(\ref{plane0})). The results are: \vspace*{2mm} \\
$F_\pi = 0.0649(68) + c_A 0.0653(25)= 0.0625(68)\quad \to \quad {\hat{F}_{\pi}} = 0.0506(55) $ ; \\
$F_K = 0.0717(45) + c_A 0.0732(23)= 0.0690(45)\quad \to \quad {\hat{F}_{K}} = 0.0568(36)$.\vspace*{2mm} \\
\noindent
We checked that the same result is obtained if we apply the method directly to ${\hat{F}_{PS}}$. Converting these results in physical units, by using $a^{-1}(m_{K^*})$, we obtain our best estimate: 
\bea
f_\pi = 139 (22)\, {\rm MeV} \quad{\rm and}\quad f_K = 156 (17)\, {\rm MeV}. 
\label{psconst}
\eea
As a consistency check, we have also applied the method to the ratios ${\hat{F}_{PS}}/M_V$, and obtained: $f_\pi = 137 (20)\, {\rm MeV}$ and $f_K = 156 (16)\, {\rm MeV}$. This kind of check is applied to other constants as well. However, it should be noted that we prefer to quote (\ref{psconst}) as our best estimate, because the lattice spacing is fixed uniquely (by $m_{K^*}$) for all the quantities considered.    

Note that, if we used the decay constants to calibrate the inverse lattice spacing we get $a^{-1}(f_\pi)=2.59(29)\, {\rm GeV}$, and $a^{-1}(f_K)=2.82(18)\, {\rm GeV}$.

As for  $SU(3)$ breaking, from the direct ratio of the decay constants, we obtain:
\bea
f_K / f_\pi - 1 = 0.123(55),
\eea
whereas  by using 
\bea
{f_K/m_{K^*}\over f_\pi/m_{\rho}} = 1.06(4)
\eea
we get  
\bea 
f_K / f_\pi - 1 = 0.136(32).
\eea
Our final estimate is then:
\bea
 f_K / f_\pi - 1 = 0.13(4),
\eea
clearly below the experimental value, $(f_K / f_\pi - 1)^{(exp)} = 0.22$. This result, nevertheless, is bigger than the prediction of one-loop quenched chiral perturbation theory: $f_K/f_\pi - 1 \simeq 0.07$ \cite{golterman}. 

\item  From the plane, $({\hat{F}}_V, M^2_{PS})$, with the same criteria as for the pseudoscalar decay constants, we extract {\underline{the vector decay constants}}:\\
${\hat{F}}_\rho = 0.0724(96),\quad $ and  $\quad {\hat{F}}_{K^*} = 0.0795(71)$,\\
which in physical units, using as before $a^{-1}(m_{K^*})$, gives:
\bea
f_\rho = 199 (15)\, {\rm MeV}, \quad{\rm and}\quad f_{K^*} = 218 (7)\, {\rm MeV}. 
\label{resvv}
\eea
This is our best estimate.
We have also applied the method to the ratios ${\hat{F}_V}/M_V$, from which we obtain: \\
$f_\rho = 199(11)\,{\rm MeV}$, and $f_{K^*} = 218(7)\,{\rm MeV}$. \\
We stress again that the reason to prefer the conservative result (\ref{resvv}), is that the lattice spacing calibrated by $m_{K^*}$
is used for both ${\hat{F}}_\rho$ and ${\hat{F}}_{K^*}$, which is clearly not the case for ${\hat{F}}_V/M_V$.

Similarly, from our data we predict \vspace*{2mm} \\
\centerline{$f_\phi / m_\phi = 0.2310(52),\,$ {\it i.e.} $f_\phi = 235(4) \, {\rm MeV}$,} \vspace*{-2mm}\\
\noindent 
For comparison, the experimental values of decay constants are (see \cite{pdg} and \cite{neubert}):\vspace*{2mm} \\
\centerline{$f_{\rho}^{(exp)} = 208(2)\,{\rm MeV}$, $\; f_{K^*}^{(exp)}=214(7)\,{\rm MeV}$ and $\; f_{\phi}^{(exp)} = 237 (4)\,{\rm MeV}$,}\vspace*{2mm}\\
\noindent 
which are obtained from 1-prong $\tau$-decays ($f_{\rho^{\pm}}$ and $f_{K^*}$), and from electromagnetic decay widths ( $f_{\rho^{0}}$ and $f_\phi$). We see that in spite of the quenched approximation, our results are in very good agreement with experimental values.

In the literature, one often encounters an alternative definition for the vector decay constant $g_V$, which is related to ours as: $g_V = f_V/M_V$. For completeness, we give the results in these units: \\
\centerline{$g_\rho = 0.260(14)$, $\; g_{K^*} = 0.246(8)$ and $\; g_{\phi} = 0.231(6)$.}\\
These values agree with Ref.~\cite{mendes}.
\noindent
We close this subsection by some useful ratios:
\bea 
{f_\rho \over f_\pi}\,=\,1.43(31),\, \quad {f_{K^*}\over f_K}\,=\,1.40(18),\, \quad {f_{K^*}\over f_\rho}\,=\,1.10(5).\nonumber
\eea
\end{itemize} 

\vspace*{.5cm}

\subsection{Coupling with tensor current}

A phenomenologically relevant quantity is the tensor coupling to the vector meson ($J^{PC} = 1^{- -}$). The values of this coupling provide the normalization of the leading-twist wave function for transversely polarized vector mesons and play an important role in light-cone sum rule analyses for heavy to light (vector) meson decays. In Ref.~\cite{braun}, a value of $f_{\rho}^{\,T}(\mu) ~=~ 160 (10)\,{\rm MeV}$ (in $\overline{\rm MS}$ scheme and at $\mu=1\, {\rm GeV}$), was obtained as best estimate, after considering different sum rule procedures intended to circumvent the problem of the pollution induced by the low lying positive parity ($J^{PC} = 1^{+ -}$) state, $b_1(1235)$, in the correlation functions relevant to the calculation. It is clearly desirable to have a lattice estimate of this quantity. We consider the same {\sl tensor-tensor} correlator as in continuum, {\it i.e.} in (\ref{eq : meff}) we take $J(x) = T_{\mu \nu}(x)= \bar{q}(x) \sigma_{\mu \nu} q(x)$. With no-spatial momentum and by choosing $T_{\mu \nu}(x)\to T_{i0}(x)$ $(i=1,2,3)$, we project out only the negative parity states (in this way, no problem of mixing with positive parity states arises), and obtain: 
\bea
\langle 0\vert {\hat{T}_{i0}}\vert V(\vec p = 0)\rangle = i e_i^{(\lambda)} {\hat{F}^{\,T}}_V M_V .
\label{eq : ten}
\eea
As in previous cases, the improvement program amounts to a redefinition\vspace*{1.5mm}\\
{\centerline{$T_{\mu \nu} (x)\to T_{\mu \nu} (x) + c_T ( \partial_\mu V_\nu (x) - \partial_\nu V_\mu (x) )$}},\vspace*{1.5mm}\\
\noindent
where the new constant $c_T$ is known perturbatively only, $c_T = 0.00896(1)C_f\, g_0^2$ \cite{a6}. With our boosted coupling $g_B = 1.256$, we have $c_T=0.0188$. Thus, the improved definition of (\ref{eq : ten}) reads:
\bea
\langle 0\vert {\hat{T}_{i0}}\vert V(\vec p = 0)\rangle = i e_i^{(\lambda)}\,M_V\, Z_T(\mu)\, ( 1\, +\, b_T\, a m_q)\, \left( F^{\,T (0)}_V \, +\, a c_T F^{\,T (1)}_V \right).
\eea
Analogously to the previous cases, out of chiral limit, we need $b_T$, where we again rely on boosted perturbation theory, {\it i.e.} $b_T=1.219$ \cite{a6}.\\
In practice we have:
\beq
 F^{\,T (0)}_V ~=~   {{\sqrt {{\cal{Z}}_{T}}}\over M_T},  \hspace*{6mm} {\rm{and}} \hspace*{6mm} 
a F^{\,T (1)}_V ~=~    - {{\sqrt {{\cal{Z}}_{V}}}\over M_V} \, {\rm  sinh}(M_V).
\eeq
The tensor current correlators were treated in the same fashion as vector ones, by averaging them over three Dirac indices. We also used the same criteria as before to fix the time-interval for fit, which in this case is slightly shorter ($t\in [13, 19]$) than the one we used for the vector correlators. 
\begin{table}
\begin{center}
\begin{tabular}{|c||c|c|c|c|c|} \hline 
{ $\kappa_1\,\kappa_2$}& { ${\ell \ell}$ }&{ $ss$}&{ $dd$}&{ $uu$}&{ critic} \\ \hline 
{\phantom{\Huge{l}}}\raisebox{-.1cm}{\phantom{\Huge{j}}}
{ $M_T$  } & 0.4914(25) & 0.4041(39) & 0.3616(58) & 0.3382(88) & 0.2833(97)  \\
{\phantom{\Huge{l}}}\raisebox{-.1cm}{\phantom{\Huge{j}}}
{ ${\cal{Z}}_T$  } & 0.0020(1) & 0.0011(1) & 0.0008(1) & 0.0006(1) & 0.0002(1)  \\ \hline
\end{tabular}
\vspace*{5mm}
\caption{\it Masses and {\rm ${\cal{Z}}$}'s of the vector mesons but with tensor current $T_{\mu \nu}(x) = \bar{q}(x) \sigma_{\mu \nu} q(x)$ (in lattice units).}
\label{tab:tensor}
\end{center}
\end{table}
\noindent
For this case, the results of our fit for the mass and the bare coupling are listed in Tab.~\ref{tab:tensor}. As expected, the results for $M_T$ are compatible with those for $M_V$, {\it i.e.} the masses obtained from the vector current correlators in Sect.2 (see Tab.~\ref{tab : mass}). Using the method of lattice planes, for the tensor couplings we obtain: \\
\bea
F_\rho^{\,T} &=& 0.0686(47) + c_T\, 0.0267(62) = 0.0691(46),\quad {\rm and}\nonumber \\
F_{K^*}^{\,T} &=& 0.0721(30) + c_T\, 0.0332(52) = 0.0728(30). 
\label{touplings}
\eea

As for the renormalization constant, contrary to the axial and vector currents, $Z_T$ is 
renormalization scheme and scale dependent. We remark, however,  that the one-loop values of $Z_T$ in Landau gauge, in RI (MOM) and in ${\overline{\rm MS}}$ schemes, are the same. We have computed the non-perturbative value of this constant in the chiral limit, in the RI (MOM) scheme, using the method of Ref.~\cite{g3}. More details will be presented in \cite{dawson}. {\begin{figure}[t!]
\vspace*{-0.2cm}
\begin{center}
\begin{tabular}{@{\hspace{0.2cm}}c c c}
  &$$\epsfbox{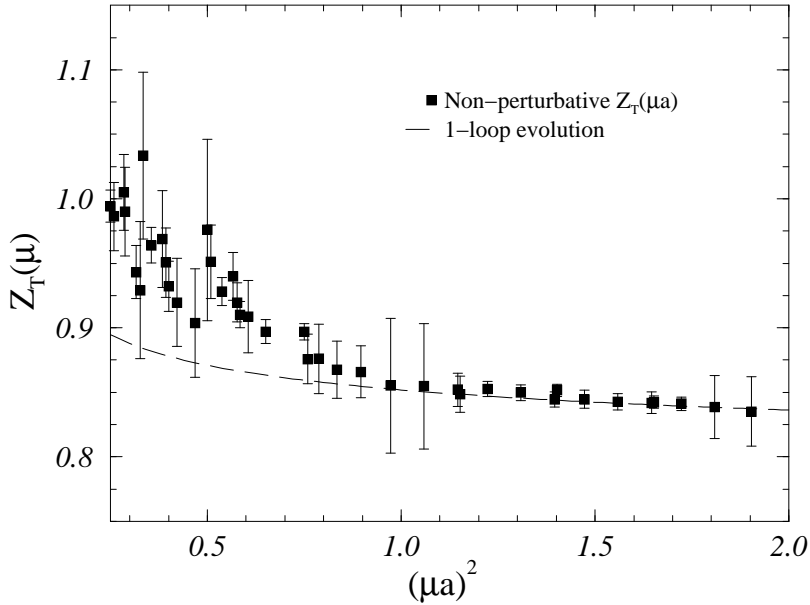}$$ & \\
\end{tabular}
\caption{\it $Z_T(\mu a)$ non-perturbatively computed on the same 100 configurations and with the same $\kappa$'s - extrapolated to the chiral limit which. Note that the one-loop evolution describes well the dependence $Z_T(\mu a)$ for $2.7\,{\rm GeV} \leq \mu \leq 3.9\,{\rm GeV}$.}
\label{zt.ps}
\end{center}
\end{figure}}
\noindent
Fig.~\ref{zt.ps}, shows that for quark virtualities between $1\leq (\mu a)^2 \leq 2$, the $one-loop$ evolution :
\bea
Z_T(\mu) = Z_T(\mu_0) \left({\alpha_s(\mu)\over \alpha_s(\mu_0)}\right)^{\gamma_0^{^{(T)}}/2 b_0}
\label{evol}
\eea
describes very well the dependence $Z_T(\mu)$ on $\mu$. The anomalous dimension of this operator is $\gamma_0^{^{(T)}}= 2 C_f$, whereas $b_0=11$ in the quenched approximation. We use (\ref{evol}) to get our non-perturbative value for $Z_T$ at $\mu=2\,{\rm GeV}$ in the MOM (or ${\overline{\rm MS}}$) scheme:
\bea
Z_T^{\rm MOM} (\mu=2\,{\rm GeV})\,=\, Z_T^{\overline{\rm MS}}(\mu=2\,{\rm GeV})\, = \,0.87(2),
\label{zteq}
\eea
where the error is mainly statistical, plus the small uncertainty in inverse lattice spacing. This result is somewhat lower than the one obtained in (boosted) perturbation theory, $Z_T(2\, {\rm GeV})=0.934(5)$ to one-loop accuracy \cite{capitani,g4} (with $c_{_{SW}}=1.614$).

\noindent
Eventually, by using the relation ${\hat{F}}_V^{\,T}(\mu) = Z_T^{\overline{\rm MS}}(\mu) ( 1+ b_T am_q) {F}_V^{\,T}$ and Eq.~(\ref{touplings}), we get the following results:
\bea
f_\rho^{\,T}(2\,{\rm GeV}) = 165(11)\, {\rm MeV},\,\,\,\,{\rm and}\,\, f_{K^*}^{\,T}(2\,{\rm GeV}) = 178(10)\, {\rm MeV} ,
\eea
where we took into account the statistical error and the error induced by the renormalization constant. This is our best estimate.
As in previous cases, we made a consistency check and calculated these constants from ${\hat{F}}_V^{\,T}(\mu)/M_T$, and obtained:\vspace*{2mm} \\
$f_\rho^{\,T}(2\,{\rm GeV}) = 161(8),$ and $f_{K^*}^{\,T}(2\,{\rm GeV}) = 178(8)\, {\rm MeV}$.\vspace*{2mm}\\
Other interesting quantities are the ratios of couplings of `different' sources of vector mesons:
\bea
\left({f_\rho^{\,T} \over f_{\rho}}\right)\hspace*{-2mm} ~=~ 0.83(13),\,\quad 
\left({f_{K^*}^{\,T} \over f_{K^*}}\right)\hspace*{-2mm} ~=~ 0.82(8).  
\eea 

\vspace*{.5cm}

\subsection{Chiral condensate}

Since we have all necessary ingredients, we can make an estimate of the value of the chiral condensate. A recent detailed discussion about different ways to extract this quantity from the lattice calculations, can be found in Ref.~\cite{vladikas}. 

First, we rely on the GMOR relation~(\ref{eq:aa}), where we use the quark mass defined by the vector~(\ref{eq:qmass1}) and/or the axial~(\ref{eq:qmass22}) Ward identities. The quark mass~(\ref{eq:qmass1}), as derived by using the vector Ward identity, was already used in the discussion on the determination of $\kappa_{crit}$.
At this point, we can combine our result for ${\hat{F}}_{_{PS}}$, with the coefficient from the fit~(\ref{eq1}), $\alpha_1=1.106(32)$. Then from~(\ref{eq:aa}), one easily obtains the relation
\bea
\langle {\bar{q}}q\rangle(\mu) = - {1\over 2} Z_{S}(\mu) {\alpha_1} {\hat{F}}_\chi^2\, 
\label{cond1}
\eea
which represents the first method~(M-I) we use to extract the chiral condensate. ${\hat{F}}_\chi$ denotes the pseudoscalar decay constant, extrapolated to the chiral limit: ${\hat{F}}_\chi = 0.0527(39)$. Equivalently, by taking the quark mass as derived from the axial Ward identity, we perform a fit analogous to the one in Eq.~(\ref{eq1}), {\it i.e.}
\bea
M_{_{PS}}^2= \tilde{\alpha}_1 (2 \rho) + \tilde{\alpha}_2 (2 \rho)^2.
\eea
The values of the parameters are:
$\tilde{\alpha}_1=1.01(7)$, and $\tilde{\alpha}_2 = 0.81(35)$. This provides us the second method~(M-II) to estimate the value of the quark condensate:
\bea
\langle {\bar{q}}q\rangle(\mu) = - {1\over 2} {Z_{P}(\mu)\over Z_A} \tilde{\alpha}_1 {\hat{F}}_\chi^2\, 
\label{cond2}
\eea
$m_q \langle\bar{q}q\rangle$ is a renormalization group invariant quantity. Taken separately, the condensate is defined only in a specific renormalization scheme and at certain scale. These details are encoded in $Z_{S,P}(\mu)$, which were recently calculated nonperturbatively \cite{dawson,vittorio}, in RI-scheme ($\rm MOM$) and in Landau gauge. The results which we use here, are obtained after extrapolation to the chiral limit: $Z_S(2\,{\rm GeV})=0.55(1)$,  $Z_P(2\,{\rm GeV})=0.43(1)$. With these values, we obtain the following results:
\bea
\langle\bar{q}q\rangle^{^{\rm (RI)}}(2\,{\rm GeV})&=& - {Z_{S}^{^{\rm (RI)}}(\mu)}\left( 1.41(31)\cdot 10^{-3} \right) = - (244\pm 25\,{\rm MeV})^3\quad {\makebox[1cm]{\rm{M-I}}}\nonumber  \\
&=& - {Z_{P}^{^{\rm (RI)}}(\mu)\over Z_A} \left( 1.30(21)\cdot 10^{-3} \right) = - (241\pm 24\,{\rm MeV})^3 \quad {\makebox[1cm]{\rm{M-II}}}
\eea
where we used $a^{-1}(m_{K^*})$ to express the result in physical units.

For the conversion of our results from $\rm MOM$ to $\overline{\rm MS}$
scheme, one relies on perturbation theory. At $\mu=2\,{\rm GeV}$, the matching was performed at the next-to-next-to-leading order in \cite{franco}. The value of the conversion factor is: $R_S^{^{NNLO}}=1.243$, and $R_S^{^{NLO}}=1.144$. Thus, in the ${\overline{\rm MS}}$-scheme, to $NLO$ accuracy, we have:\\
\bea
\langle\bar{q}q\rangle^{^{\overline{\rm MS}}} (2\,{\rm GeV})&=& - (255\pm 26\,{\rm MeV})^3_{NLO}\quad  {\makebox[1cm]{\rm{M-I}}}\nonumber \\
&=& - (252\pm 25\,{\rm MeV})^3_{NLO}\quad  {\makebox[1cm]{\rm{M-II}}} .
\label{cond3}
\eea

Another possibility~(M-III) to estimate the value of the chiral condensate is provided by the low energy theorem:
\bea
 \langle 0 \vert \bar{q}i \gamma_5 q\vert \pi \rangle \, =\, - {2\over f_\pi}\langle 0 \vert \bar{q} q\vert 0 \rangle\, . 
\label{low}
\eea
For this purpose, we take the results for the matrix element $\vert\langle 0 \vert P(0) \vert PS \rangle \vert^2= {\cal{Z}}_{PS}$, listed in Tab.~(\ref{tab : mass}). With its value extrapolated to the chiral limit, ${\cal{Z}}_\chi$, as well as with $\hat{F}_\chi$, Eq.~(\ref{low}) amounts to the following:

\bea
\langle \bar{q}q \rangle^{^{\rm (RI)}}(2\,{\rm GeV}) &=& - {Z_{P}^{^{\rm (RI)}}(\mu)} {\hat{F}_\chi \over 2} \sqrt{{\cal{Z}}_\chi} = -( 241\pm 20\,{\rm MeV})^3 \nonumber \cr
 \to & & \langle\bar{q}q\rangle^{^{\overline{\rm MS}}} (2\,{\rm GeV}) = -( 252\pm 21\,{\rm MeV})^3_{_{NLO}}\quad  {\makebox[1cm]{\rm{M-III}}} .
\eea
Altogether, we quote 
\bea
\langle\bar{q}q\rangle^{^{\overline{\rm MS}}} (2\,{\rm GeV}) = -( 253\pm 25\,{\rm MeV})^3_{_{NLO}}, 
\eea
as our final result{\footnote{As indicated, all our estimates of $\langle\bar{q}q\rangle^{^{\overline{\rm MS}}} (2\,{\rm GeV})$ are given at NLO accuracy, so that it is easier to make a comparison with results of other approaches and other lattice groups. The results to NNLO can be obtained trivially, by multiplication by $R_S^{^{NNLO}}/R_S^{^{NLO}}$.}}. Our estimates agree with results of Ref.~\cite{vladikas}.  

\noindent
Note also, that by using the evolution equation to $NLO$ (with $\Lambda^{(4)}=300\,{\rm MeV}$), we obtain:
$\langle\bar{q}q\rangle^{^{\overline{\rm MS}}}_{_{NLO}} (1\,{\rm GeV})= - ( 232\pm 24\,{\rm MeV})^3$,
which is very close to the value suggested by authors of Ref.~\cite{leutwyler},   $\langle\bar{q}q\rangle^{^{\overline{\rm MS}}}(1\,{\rm GeV})\simeq - ( 225 \pm 25 \,{\rm MeV})^3$. 

\section{Comparison with unimproved results}
\setcounter{equation}{0}

\begin{table}[hbt]
\begin{center}
\begin{tabular}{c|ccc}
{\rm Action}& \multicolumn{3}{c}{\bf Wilson ($c_{_{SW}}=0$)}   \\
\hline 
{\phantom{\Huge{l}}}\raisebox{-.1cm}{\phantom{\Huge{j}}}
$\beta$ & 6.0 & 6.2& 6.4\\ 
{\phantom{\large{l}}}\raisebox{-.1cm}{\phantom{\large{j}}}
{\rm \# Confs}&320&250 &400\\
{\phantom{\large{l}}}\raisebox{-.1cm}{\phantom{\large{j}}}
{\rm Volume}  &$18^3\times 64$&$24^3\times 64$&$24^3\times 64$ \\
\hline
{\phantom{\Huge{l}}}\raisebox{-.1cm}{\phantom{\Huge{j}}}
$a^{-1}(m_\rho) [{\rm GeV}]$
            &2.14(5) &2.84(11) & 4.00(17)  \\
{\phantom{\large{l}}}\raisebox{-.1cm}{\phantom{\large{j}}}
$a^{-1}(m_{K^*}) [{\rm GeV}]$          
           &2.26(5)&3.00(9)&4.15(16) \\
\end{tabular}
\caption{\it Summary of lattice characteristics used for the comparison with the. $a^{-1}(m_\rho)$ is obtained by the conventional - and $a^{-1}(m_{K^*})$ by the lattice plane method.} 
\label{tab : razne}
\end{center}
\end{table}

In this section, we make a `consistent' comparison of our results with the unimproved data, which were generated by the ${\sc APE}$ collaboration in previous studies using the Wilson action. By `consistent', we mean that the same methods to extract the physical quantities were used both in the improved and unimproved cases. In Tab.~\ref{tab : razne}, we give some basic information on the simulation with the Wilson action. We refer the reader to Ref.~\cite{giusti} for more details.

The physical volume for $\beta=6.0$ and $\beta=6.2$ is approximately constant {$\simeq\,(1.7\,{\rm fm})^3$}, whereas for $\beta=6.4$ it is somewhat smaller {$\simeq\,(1.2\,{\rm fm})^3$}. Since all criteria used in Sec.~2.,3. and 4. to extract masses and decay constants are applied to all lattice data, we can reliably investigate the effects of improvement. 
To extract physical observables, we employ the lattice plane method, since no explicit extrapolation to $\kappa_{crit}$ is needed. The scale for each lattice is fixed by $m_{K^*}$ and the results which are to be compared, are listed in Tab.~\ref{tab : razneplane}.

\begin{table}[h!]
\begin{center}
\begin{tabular}{|c|ccc|c|}
\hline 
{\phantom{\Huge{l}}}\raisebox{-.2cm}{\phantom{\Huge{j}}}
{\sl Action}& \multicolumn{3}{c|}{\bf Wilson }   & \multicolumn{1}{c|}{\bf Clover - NP } \\
\hline 
{\phantom{\Huge{l}}}\raisebox{-.1cm}{\phantom{\Huge{j}}}
$\beta$ & 6.0 & 6.2& 6.4 & 6.2\\ 
$a(m_{K^*})\, [{\rm fm}]$
            &0.087(2) &0.066(2) & 0.048(2)  & 0.072(3)\\
\hline 
{\phantom{\Huge{l}}}\raisebox{-.1cm}{\phantom{\Huge{j}}}
$f_{\pi}\, [{\rm GeV}]$
            &0.160(7) &0.138(7) & 0.150(10) & 0.139(22)\\
{\phantom{\Huge{l}}}\raisebox{-.1cm}{\phantom{\Huge{j}}}
$f_{K}\, [{\rm GeV}]$
            &0.172(6) &0.155(5) & 0.164(8) & 0.156(17)\\
{\phantom{\Huge{l}}}\raisebox{-.1cm}{\phantom{\Huge{j}}}
$(f_{K}/f_{\pi}) - 1$
            &0.076(10) &0.126(14) & 0.095(16) & 0.123(55)\\
{\phantom{\Huge{l}}}\raisebox{-.1cm}{\phantom{\Huge{j}}}
$f_{\rho}\, [{\rm GeV}]$
            &0.310(14) &0.286(16) & 0.241(12) & 0.199(15)\\
{\phantom{\Huge{l}}}\raisebox{-.1cm}{\phantom{\Huge{j}}}
$f_{K^*}\, [{\rm GeV}]$
            &0.312(10) &0.291(10) & 0.255(8) & 0.218(7)\\
\hline
\end{tabular}
\caption{\it Comparison of several physical quantities calculated using Wilson fermions: nonimproved and non-perturbatively improved. The lattice scale was fixed by $m_{K^*}$.}
\label{tab : razneplane}
\end{center}
\end{table}
{\begin{figure}[h!!]
\vspace*{-1.8cm}
\begin{center}
\begin{tabular}{@{\hspace{-1.8cm}}c c c}
$$\epsfbox{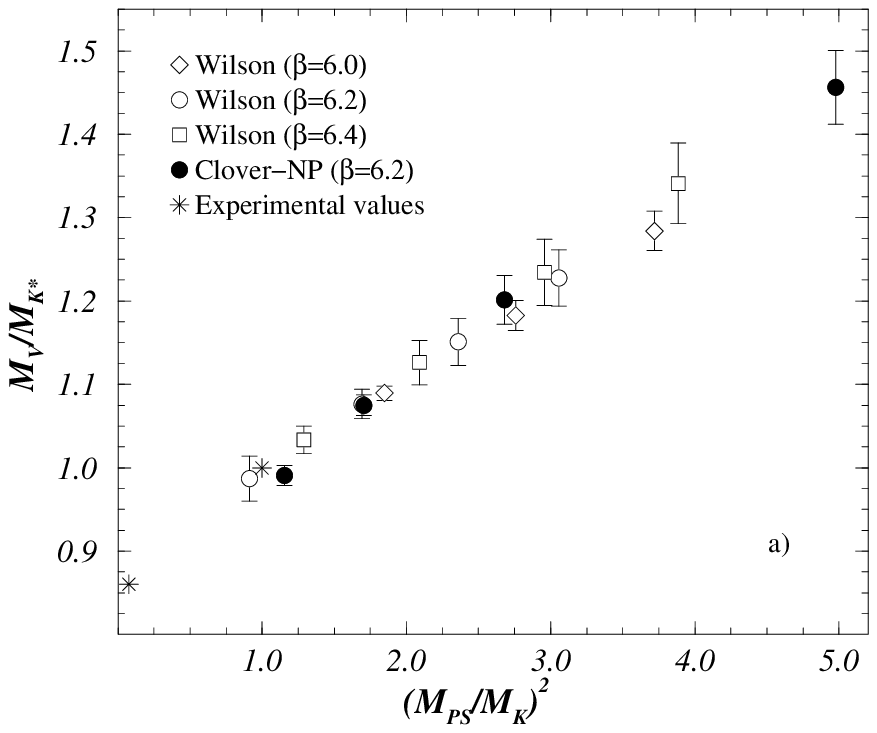}$$& {\hspace{-2.2cm}}
  &$$\epsfbox{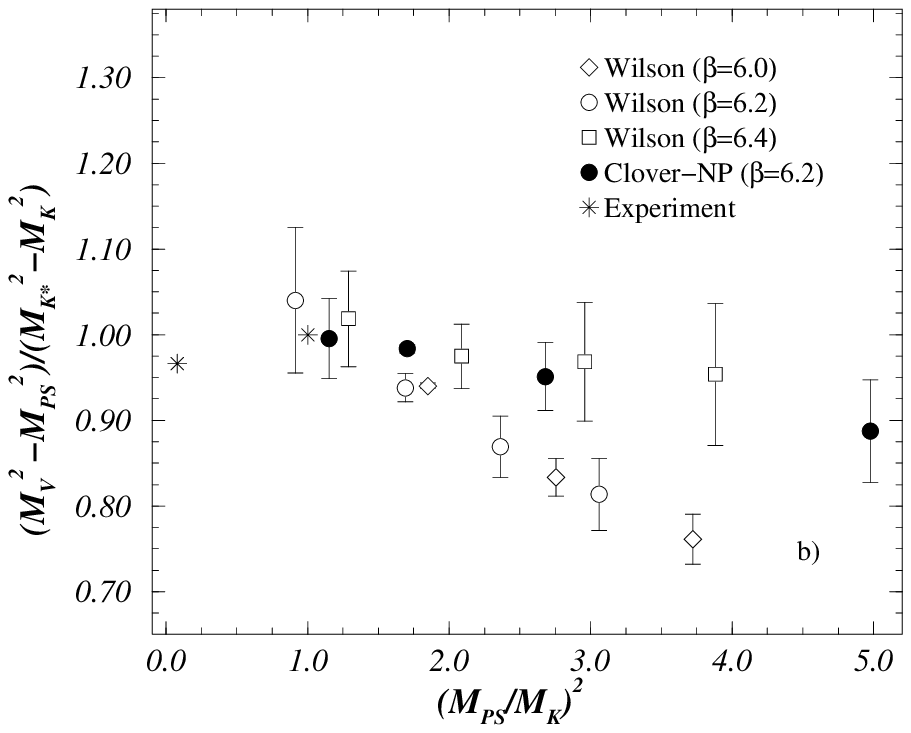}$$  \\
$$\epsfbox{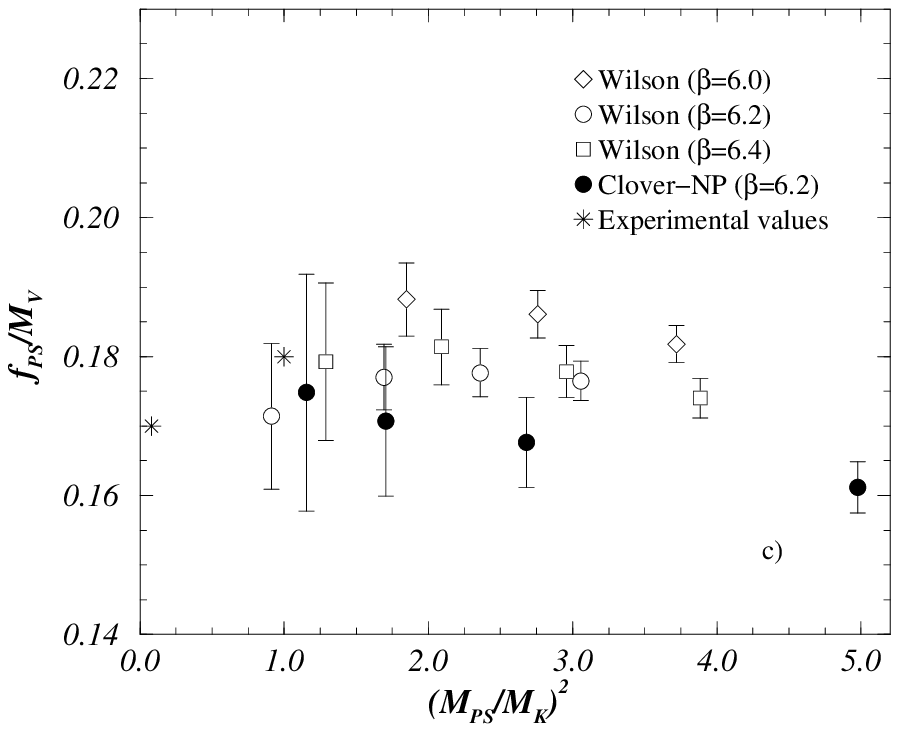}$$& {\hspace{-2.2cm}}
  &$$\epsfbox{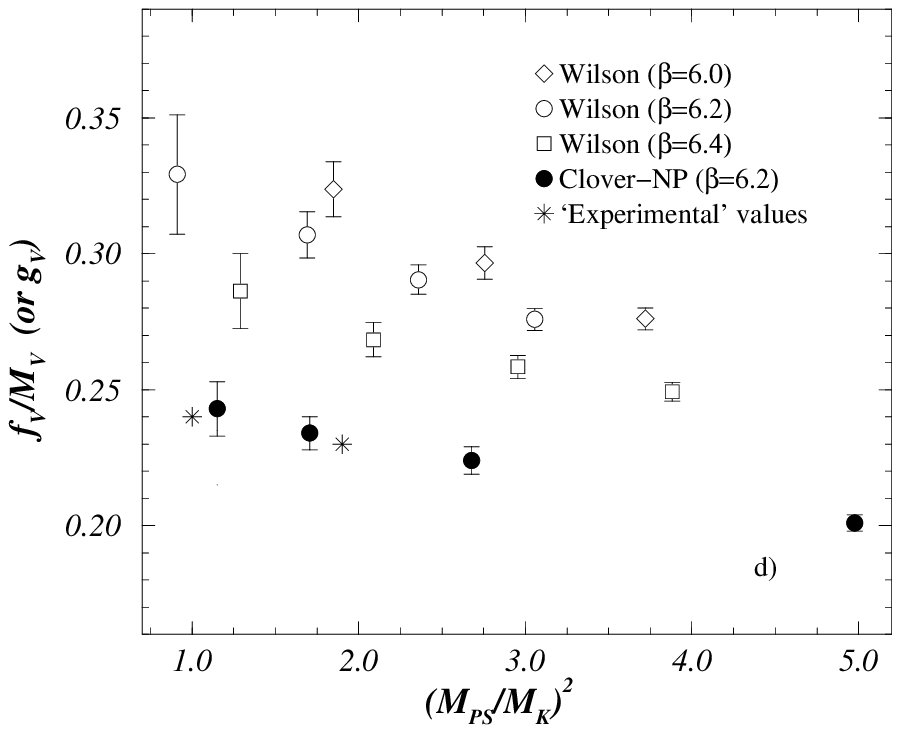}$$  \\
\end{tabular}
\vspace*{7mm}

\caption{\it Comparison of improved (Clover -NP) with unimproved (Wilson) results. {\rm a)} and particularly {\rm b)} show the effect of improvement for the spectrum. {\rm c)} and {\rm d)} show this effect for decay constants.}
\label{fig:razne1}
\end{center}
\vspace*{-0.6cm}
\end{figure}}
\noindent

To have some better insight on the effect of the improvement, we plot in Figs.~\ref{fig:razne1} four dimensionless ratios. 
We constructed these ratios using quantities directly extracted from correlation functions. From Figs.~\ref{fig:razne1}a,b we see that the mass dependence of our results is very close to that obtained with the Wilson (unimproved) action at $\beta=6.4$. However, this conclusion is only qualitative since the physical volume used for calculations at $\beta=6.4$ was small.  

As for the decay constants (Fig.~\ref{fig:razne1}c,d), we use in both cases currents which are non-perturbatively renormalized. So the differences have to be attributed to genuine ${\cal{O}}(a)$ effect.
In the unimproved case, a symptom of the presence of large ${\cal{O}}(a)$ effects was particularly evident in the vector meson decay constant. Particularly problematic was the determination of $Z_V$ \cite{abada}. Very recently in Ref.~\cite{fred}, $Z_V$ was calculated by using the program for non-perturbative renormalization~\cite{g3}, but the values for the vector decay constants remained well above the experimental ones. We compare our results in Figs.~\ref{fig:razne1}c,d. We notice a significant change for the vector decay constant, less pronounced for the pseudoscalar one. This change improves the agreement of the quenched lattice results with the experimental values. We reiterate that the improved results would be about $5\%$ bigger, had we used the value of $c_V$ from boosted perturbation theory. The reason to point this out is that the nonperturbatively determined $c_V$ is almost by an order of magnitude bigger than the corresponding perturbative value. While this observation does not substantially alter our results in the light meson sector of light meson, it seriously affects the determination of the heavy-light vector meson decay constants~\cite{prepa}.

To better monitor the effect of the improvement, we also make the comparison of the vector decay constants with the results obtained by using the tree-level improved Clover action ($c_{_{SW}}=1$), at the same $\beta=6.2$, and the same volume $24^3\times 64$ ({\sl see} \cite{giusti} {\sl for details}). In Fig.~{$7$}, we observe that the effect of the tree-level improvement is a slight decrease of $f_V$. Further decrease towards the experimental values is the effect of the full elimination of ${\cal{O}}(a)$ errors. 


\begin{figure}[h!]
\begin{center}
\begin{tabular}{@{\hspace{-0.7cm}}c c c}
  &$$\epsfbox{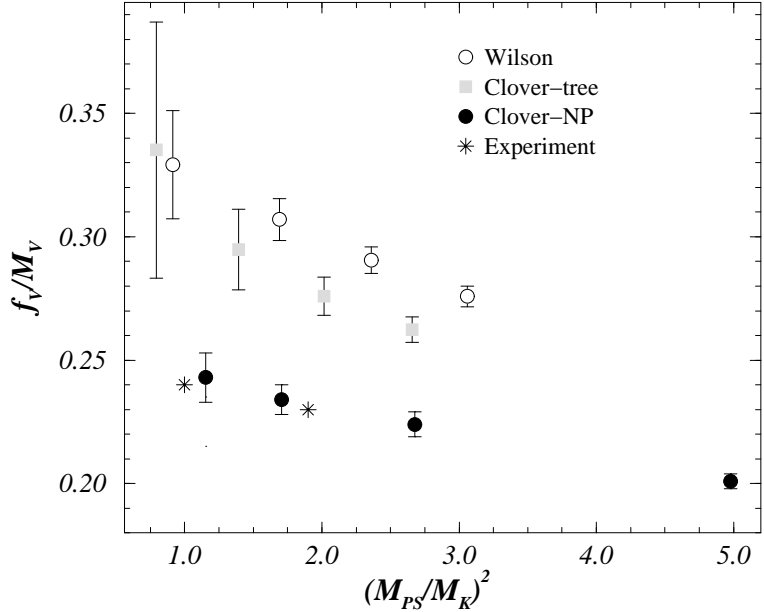}$$ & \\
\end{tabular}
{\caption{\it The effect of improvement on the vector meson decay constants. All results are obtained at $\beta=6.2$ and the volume $24^3\times 64$, with unimproved $(c_{_{SW}}=0)$, tree-level improved $(c_{_{SW}}=1)$, and non-perturbatively $(c_{_{SW}}=1.614)$ improved Wilson action which are denoted as Wilson, Clover-tree and Clover-NP, respectively.}}	

\label{cloverrr}

\end{center}
\end{figure}

\vspace{1cm}

\section{Energy momentum relation}
\setcounter{equation}{0}

For studies of semileptonic and radiative decays on the lattice, it is important to calculate the form factors as functions of the momentum  transfer. This is achieved by giving different momenta to the interacting hadrons. The injection of momentum introduces further discretization errors, in particular, affecting the continuum dispersion relation: 
\bea
E(\vec{P})^2 = \vec{P}^2 + M^2. 
\label{cont0}
\eea
In order to investigate lattice artifacts related to this problem, we studied the meson propagators at several values of $\vec{P}$. For this purpose, we used the Eq.(\ref{cont0}), as well as the lattice free boson dispersion relation:
\bea
{\rm sinh}^2 \left({E(\vec{P})\over 2}\right) = {\rm sinh}^2 \left({M \over 2}\right) + \sum_{i=1}^{3} {\rm sin}^2 \left({\vec{P} \over 2}\right)
\label{eq:dr}
\eea
which may be derived from the discretized free boson action with  nearest neighbors interaction.
This is not a unique choice since it depends on the way we define derivatives on the lattice. In previous studies \cite{giusti,gupta1}, the authors have shown that the choice (\ref{eq:dr}) is indeed favored by data.

We used lattice cubic symmetry, parity and charge conjugation to relate symmetric configurations and thus increase the statistical quality of our correlators. By writing ${\vec{P}} \equiv  2 \pi / L a \, ( n_x, n_y, n_z)$, the components of spatial momenta for pseudoscalar and vector correlation functions that we consider here, are:
\bea
(0,0,0);\,(1,0,0);\,(1,1,0);\,(1,1,1);\nonumber \\
(2,0,0);\,(2,1,0);\,(2,1,1);\,(2,2,0).
\eea
We limited ourselves to $|\vec{P}|^2 \leq 8$, after a test-run where we observed that for higher momenta, the hadronic propagators were immediately drowning into very large noise. We used the same criteria as in Sec.~2 to establish the time intervals for the fit at each momentum considered. These time intervals are listed in Tab.~\ref{times1}. Note that for large momenta, the fit intervals become very short.
\begin{table}[h!]
\begin{center}
\begin{tabular}{c||c|c|c|c|c|c|c} 
{ $|{\vec P}|$}& { $1$} & { $\sqrt{2}$} & { $\sqrt{3}$} & { $2$} & { $\sqrt{5}$} & { $\sqrt{6}$} & { $\sqrt{8}$} \\  \hline
{ ${\rm PS:}\,t_{min}-t_{max}$}& { 10-23} & { 10-19} & {10-18} & {10-15} & {10-14} & {10-13} & {10-12} \\  
{ ${\rm VEC:}\,t_{min}-t_{max}$}& { 11-24} & {11-21} & {11-18} & { 11-18} & {11-17} & { 11-15} & {11-14} 
\end{tabular}
\caption{\it Values of the times for fit for pseudoscalar and vector correlation functions for all momenta considered in this work.}
\label{times1}
\end{center}
\end{table}
{
\begin{table}[h!!]
\vspace*{-0.2cm}
{
\begin{center}
\begin{tabular}{||c|c|c|c|c||} 
\hline \hline 

{\phantom{\huge{l}}}\raisebox{-.2cm}{\phantom{\huge{j}}}
{ }& { $|\vec P|^2=0$} & { $|\vec P|^2=1$}& { $|\vec P|^2=2$} &  { $|\vec P|^2=3$} \\ \hline 

{$\rm vector$ {\it{ll}}}& 0.4911(29) &0.5576(24)&0.6143(22)&0.6680(21)\\

{$\rm pseudoscalar$ {\it{ll}}} & 0.4167(15)&0.4957(18)&0.5520(20)&0.6190(23) \\

{$\rm vector$  {\it{ss}}}& 0.4055(47) &0.4839(36)&0.5478(33)&0.6099(33) \\

{$\rm pseudoscalar$ {\it{ss}}} & 0.3058(19)&0.4041(22)&0.4829(24)&0.5493(34) \\ 

{$\rm vector$  {\it{dd}}}& 0.3626(78) &0.4493(58)&0.5168(49)&0.5838(49) \\

{$\rm pseudoscalar$ {\it{dd}}}& 0.2440(21) &0.3558(27)&0.4409(32)&0.5077(49)) \\ 

{$\rm vector$  {\it{uu}}} & 0.335(12)&0.4281(74)&0.4976(69)&0.5678(67) \\

{$\rm pseudoscalar$ {\it{uu}}}& 0.2007(26) &0.3242(40)&0.4156(41)&0.4817(70) \\ 
\hline
{\phantom{\huge{l}}}\raisebox{-.2cm}{\phantom{\huge{j}}}
{ }&{ $|\vec P|^2=4$} & { $|\vec P|^2=5$} &  { $|\vec P|^2=6$} &  { $|\vec P|^2=8$}\\ \hline 
{$\rm vector$ {\it{ll}}}& 0.7084(22)&0.7518(22)&0.7933(33)&0.8623(35) \\
{$\rm pseudoscalar$ {\it{ll}}}&0.6678(28) &0.7138(34)& 0.7556(48)&0.828(7)\\
{$\rm vector$  {\it{ss}}}&0.6515(32) &0.6987(33)&0.7454(50)&0.8165(45)\\
{$\rm pseudoscalar$ {\it{ss}}}&0.6029(46)&0.6527(52)&0.6989(83)&0.779(15)\\ 
{$\rm vector$  {\it{dd}}}&0.6253(32) &0.6729(44)&0.7225(71)&0.7931(56)\\
{$\rm pseudoscalar$ {\it{dd}}}&0.5720(79)&0.6198(82)&0.6712(139)&0.767(31)\\ 
{$\rm vector$  {\it{uu}}}&0.6102(57) &0.6572(55)&0.7116(89) & 0.7783(67)\\
{$\rm pseudoscalar$ {\it{uu}}}&0.5633(152)&0.6029(108)&0.6595(220)&0.779(65)\\ 
\hline \hline
\end{tabular}
\vspace*{1cm}
\caption{\it Values of energies $E(\vec P)$, as extracted from the fit to our data for different spatial momentum injections and for fixed{ \rm {${\cal{Z}} = {\cal{Z}}({\vec P} = 0)$})}.}
\label{kkk}
\end{center}
}
\end{table}
}
{
\begin{figure}
\vspace*{-1.2cm}
\begin{center}
\begin{tabular}{@{\hspace{-1.5cm}}c c c }
\hfill &$$\epsfbox{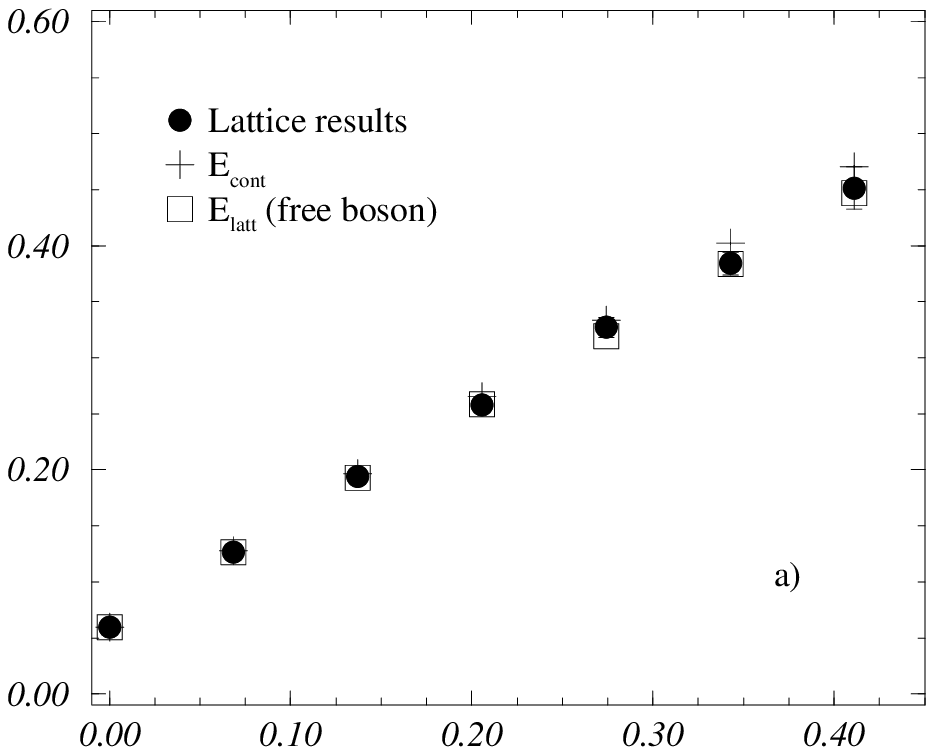}$$ &\hfill  \\
\vspace*{.2cm}
\hfill &$$\epsfbox{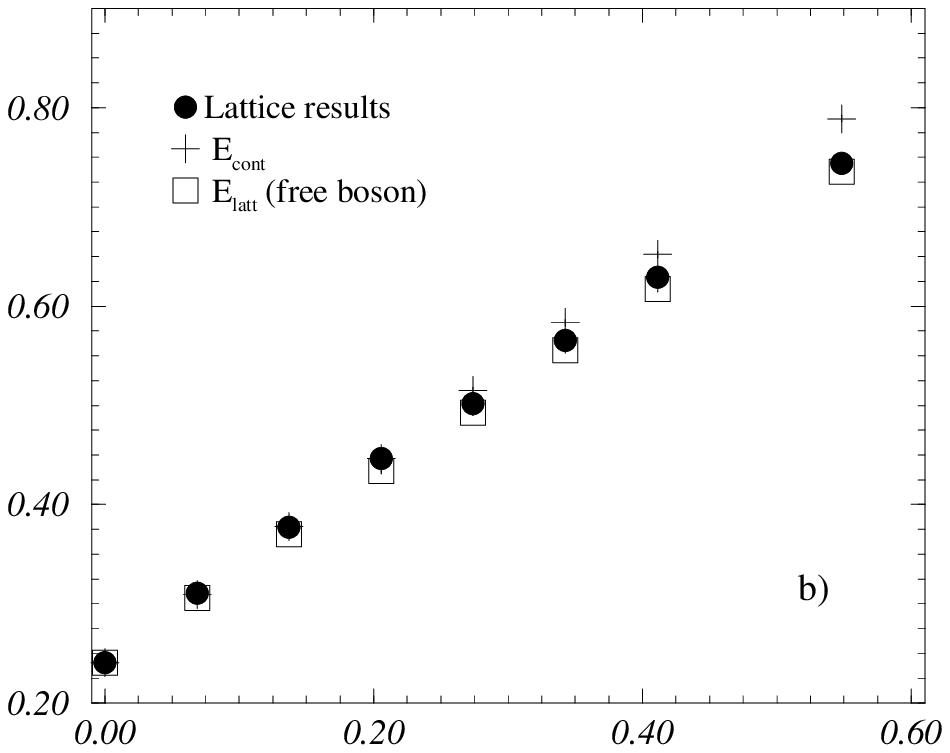}$$ &\hfill  \\
\end{tabular}
\caption{\it Energy-momentum relation for light-pseudoscalar {\rm (a)} and vector {\rm (b)} mesons: In both figures $E^2$ vs. $|\vec{P}|^2$ as obtained from the fit to {\rm(\ref{fit6})}, are denoted by bullets; {\rm [$+$]} mark values obtained by {\rm (\ref{cont0})} , while {\rm [$\Box$]} denote values obtained by using {\rm (\ref{eq:dr})}, where the masses are listed in {\rm Tab.~\ref{tab : mass}}.}
\label{rel9}
\end{center}
\end{figure}
}
\newline
\noindent
To check the relations (\ref{cont0}) and (\ref{eq:dr}), we fit our data either with~
\bea
C_{JJ}(t,{\vec{P}}) = {{\cal{Z}}_J\over 2~ E({\vec{P}})} e^{-E({\vec{P}}) t},
\label{fit66}
\eea
\noindent
or with
\bea
C_{JJ}(t,{\vec{P}}) = {{\cal{Z}}_J\over 2~{\rm sinh}\left( E({\vec{P}})\right)} e^{-E({\vec{P}}) t}.
\label{fit6}
\eea
The same relations hold for transversely polarized vector mesons which we consider here. In general, however, the term due to the polarization should be accounted for. 

On the basis of the Wigner-Eckart theorem, we fix ${\cal{Z}}_J$ at $|\vec{P}|=0$, to fit $E(\vec{P})$. We checked and realized that the relation~(\ref{fit6}) provides a much better fit to our data, starting from the three-momenta $\vert \vec{P}\vert^2\geq 3$. By using (\ref{fit6}),
we find that the agreement with (\ref{eq:dr}) is excellent, up to rather large values of injected momenta. The results of these fits are given in Tab.~\ref{kkk}, and illustrated in Fig.~\ref{rel9}. In that figure, we depict our data by `bullets' describing the results obtained from the fit to (\ref{fit6}). For better comparison with dispersion relations, we also show $E(\vec{P})$ as obtained from (\ref{cont0}) and (\ref{eq:dr}) and represent them by `plus' and `square' symbols respectively. The mass terms in these dispersion relation are those extracted from the correlators with $|\vec{P}|=0$, {\it i.e.} those which we already listed in Tab.~\ref{tab : mass}. A comparison with previous results is somewhat difficult since the simulations with the Wilson action were done for $|\vec{P}|^2 \leq 4$ only. This makes it difficult to distinguish which dispersion relation is better to use. It is important to note that our data are also compatible with both dispersion relations, (\ref{cont0}) and (\ref{eq:dr}), when the small momenta are considered. Only for large momenta it becomes clear that the Eq.~(\ref{eq:dr}) describes our data much better. 

\clearpage

\section{Conclusion}
\setcounter{equation}{0}
In conclusion, we summarize the main results of this analysis. We performed a lattice study of the light hadron spectrum and meson decay constants with non-perturbatively improved action and operators. We showed that, for directly accessible (not so light) quark masses, the physical contribution of quadratic quark mass terms to the mass of the pseudoscalar mesons, exceeds the effect of lattice artifacts. Thus we conclude that, in the quenched case, there is a positive contribution to $m_{PS}^2$ in $m_q^2$. For all the quantities considered here, we determined the physical values using the method of physical lattice planes. Our main results are the following:
\begin{itemize}
\item[--] In spite of the use of the quenched approximation, we find that the values of inverse lattice spacing, as obtained from different physical quantities,
\begin{table}[h!!]
\vspace*{-6mm}
\begin{center}
\begin{tabular}{c|c c c c} 
\multicolumn{1}{c|}{\rm Ref.~\cite{bali}} & \multicolumn{4}{c}{\rm this work}\\ \hline
${a^{-1}(\sigma)}$ & ${a^{-1}(m_{K^*})}$ & ${a^{-1}(m_\rho)}$  & ${a^{-1}(f_K)}$ & ${a^{-1}(f_\pi)}$ \\ 
{ $2.72(3)\,{\rm GeV}$}&{ $2.75(17)\,{\rm GeV}$}&{ $2.75(22)\,{\rm GeV}$}&{ $2.82(18)\,{\rm GeV}$}&{ $2.59(29)\,{\rm GeV}$}
\end{tabular}
\label{times}
\end{center}
\vspace*{-6mm}
\end{table}
are compatible with each other (within the errors), and with the one obtained from the string tension{\footnote{It is worth mentioning that the inverse lattice spacing in Ref.~\cite{bali}, was obtained by using $\sqrt{\sigma}=440\, {\rm MeV}$.}}. Since the least extrapolation is needed for $M_{K^*}$, we decided to fix the scale by this quantity. 
\item[--] We verified that the hyperfine splitting in the region of light mesons is well reproduced by our data. 
We have also extracted $J=0.37$, but noticed that by using the quadratic fit $M_V(M_{PS}^2)$, one gets $J=0.47(6)$
which is (despite its large error) closer to the experimental value. This point deserves further investigation.
 
\item[--] For the pseudoscalar decay constants, we have:\\ $f_K=156\pm 17\,{\rm MeV}$, $f_\pi =139\pm 22\,{\rm MeV}$, whereas the SU(3) breaking is (in spite of improvement) smaller than the experiment, {\it i.e.} $(f_K - f_\pi)/f_\pi = 0.13(4)$.

\item[--] For the vector decay constants, we obtain:\\ $f_{K^*}=219\pm 7\,{\rm MeV}$, $f_\rho =199\pm 15\,{\rm MeV}$, $f_\phi =235\pm 4\,{\rm MeV}$.
\item[--] In determination of the coupling of the vector meson to the  tensor current, we also calculated $Z_T(\mu)$ non-perturbatively, which is a new result. In Landau gauge, and at $\mu=2\,{\rm GeV}$, we extracted:\\
$Z_T^{\overline{\rm MS}}=Z_T^{\rm MOM}= 0.87(2).$\\
Results for tensor couplings are:\\
$ f_{K^*}^{^T}(2\,{\rm GeV}) = 178\pm 10\,{\rm MeV}$, $f_\rho^{^T}(2\,{\rm GeV}) =165\pm 11\,{\rm MeV}$.
\item[--] By using the Gell-Mann-Oakes-Renner formula, we estimated the value of the chiral condensate:\\
$\langle\bar{q}q\rangle^{^{\overline{\rm MS}}} (2\,{\rm GeV})= - ( 253\pm 25\,{\rm MeV})^3$.
\item[--] By using the same criteria of analysis, we consistently compared ours to the results obtained with unimproved Wilson fermions. We conclude the following:
\begin{itemize}
\item[o] Our meson masses and hyperfine splitting, as directly extracted from correlation functions, are qualitatively comparable to those obtained without improvement at $\beta=6.4$,  
\item[o] The contribution of ${\cal{O}}(a)$ term in improved operator is of order of $5\%$. A striking effect of the full ${\cal{O}}(a)$ non-perturbative improvement is evident in the case of the decay constant for vector mesons, and less so for the pseudoscalars.
\end{itemize}   
\item[--] We explored a wide range of momentum injections to study the energy-momentum relation for pseudoscalar and vector mesons. The lattice artifacts become important for higher momenta, but we show that the lattice dispersion relation for a free boson fits our data to excellent accuracy. 
This conclusion will be very useful for the study of semileptonic decays of heavy to light mesons.  
\end{itemize}

\vspace*{1.6cm}

\noindent
{\large{Acknowledgement}}

\vspace*{10mm}

D.B. takes pleasure to thank ``Fondation des Treilles'' for financial support. V.L. and G.M. acknowledge the M.U.R.S.T. and the INFN for partial support. 

\vspace*{20mm}

\end{document}